\documentclass[12pt,a4paper]{article}
\usepackage{graphicx}% Include figure files
\usepackage{dcolumn}% Align table columns on decimal point
\usepackage{bm}% bold math
\usepackage{amsmath,amsfonts,amssymb,mathrsfs}
\usepackage[dvipsnames]{xcolor}
\usepackage{bbm}
\usepackage{slashed, MnSymbol}
\usepackage{braket,xcolor}
\usepackage{verbatim}
\usepackage{subcaption}
\usepackage[normalem]{ulem}
\usepackage{multirow}
\usepackage{amsfonts,mathtools,resizegather}
\usepackage[normalem]{ulem}
\usepackage[utf8]{inputenc}
\usepackage{scalerel}
\usepackage{makecell}
\usepackage{dcolumn}
\usepackage{amsfonts, mathtools,resizegather}
\usepackage{multicol}
\usepackage{subcaption}
\usepackage[dvipsnames]{xcolor}
\usepackage[bb=boondox]{mathalfa}
\usepackage[export]{adjustbox}
\usepackage[normalem]{ulem}
\usepackage{caption,jheppub}

\DeclareMathOperator{\sech}{sech}

\DeclareMathOperator{\arcsinh}{arcsinh}
\newcommand{\psis}{\psi_{i_{1}}\psi_{i_{2}}\cdots\psi_{i_{s}}}
\newcommand{\psial}{\psi_{\alpha_{1}}\psi_{\alpha_{2}}\cdots\psi_{\alpha_{p}}}
\newcommand{\psibe}{\psi_{\beta_{1}}\psi_{\beta_{2}}\cdots\psi_{\beta_{p}}}
\newcommand{\val}{V^{k}_{\alpha_1 \alpha_2 \dots \alpha_p}}
\newcommand{\ovral}{\overline{V}^{k}_{\beta_1 \beta_2 \dots \beta_p}}
\newcommand{\vbe}{V^{k}_{\beta_1 \beta_2 \dots \beta_p}}

 \title{\boldmath {Operator dynamics in Lindbladian SYK: a Krylov complexity perspective}}

\author[a]{Budhaditya Bhattacharjee,}
\author[b,c]{Pratik Nandy,}
\author[d]{and Tanay Pathak}
\affiliation[a]{Center for Theoretical Physics of Complex Systems,
Institute for Basic Science,\\ Daejeon - 34126,  Republic of Korea}
\affiliation[b]{Center for Gravitational Physics and Quantum Information,\\
Yukawa Institute for Theoretical Physics, Kyoto University,\\
Kitashirakawa Oiwakecho, Sakyo-ku, Kyoto 606-8502, Japan}
\affiliation[c]{RIKEN Interdisciplinary Theoretical and Mathematical Sciences Program (iTHEMS),\\
Wako, Saitama 351-0198, Japan}
\affiliation[d]{Centre for High Energy Physics, Indian Institute of Science,\\ C.V. Raman Avenue, Bangalore 560012, India}

\emailAdd{budhadityab@ibs.re.kr}
\emailAdd{pratik@yukawa.kyoto-u.ac.jp}
\emailAdd{tanaypathak@iisc.ac.in}

\abstract{We use Krylov complexity to study operator growth in the $q$-body dissipative Sachdev-Ye-Kitaev (SYK) model, where the dissipation is modeled by linear and random $p$-body Lindblad operators. In the large $q$ limit, we analytically establish the linear growth of two sets of coefficients for any generic jump operators. We numerically verify this by implementing the bi-Lanczos algorithm, which transforms the Lindbladian into a pure tridiagonal form. We find that the Krylov complexity saturates inversely with the dissipation strength, while the dissipative timescale grows logarithmically. This is akin to the behavior of other $\mathfrak{q}$-complexity measures, namely out-of-time-order correlator (OTOC) and operator size, which we also demonstrate. We connect these observations to continuous quantum measurement processes. We further investigate the pole structure of a generic auto-correlation and the high-frequency behavior of the spectral function in the presence of dissipation, thereby revealing a general principle for operator growth in dissipative quantum chaotic systems.}

\begin{document}
\maketitle
\flushbottom

\section{Introduction}
Operator growth is a useful way of distinguishing quantum integrable systems from those exhibiting quantum chaos and information scrambling. It probes how fast a local operator grows under the time evolution by the Hamiltonian of the system. Several ways of characterizing such growth have been proposed in recent years, namely the operator size distribution \cite{Roberts:2014isa, Roberts:2018mnp}, out-of-time-ordered correlator (OTOC) \cite{Maldacena:2015waa}, and Krylov complexity \cite{Parker:2018yvk}. These approaches are indirect since they require a probe operator and contrast with more direct probes such as level statistics \cite{BGS, Wigner} and spectral form factor (SFF) \cite{Cotler:2016fpe}. The latter is particularly useful in the semiclassical theory of gravity, where a dip, ramp, and plateau are observed and feature behavior similar to underlying random matrix universality at late times. On the other hand operator growth turns out to be useful in probing the black hole interior. Particularly the increasing momentum of a particle in a near AdS$_2$ black hole is reflected as an operator growth in the dual geometry \cite{Brown:2018kvn}, often studied under the shadow of operator complexity \cite{Qi:2018bje, Kar:2021nbm}.

In this paper, our interest is in studying Krylov complexity, which measures operator growth on a special basis known as the Krylov basis. The basis is formed by an iterative Gram-Schmidt-like orthonormalization procedure known as the Lanczos algorithm \cite{Lanczos1950AnIM,viswanath1994recursion}. An output of this algorithm is the Lanczos coefficients, which usually show distinctive features for integrable and non-integrable systems, albeit in some special cases \cite{Dymarsky:2021bjq, Bhattacharjee:2022vlt}. The universal operator growth hypothesis, proposed in \cite{Parker:2018yvk} states that the Lanczos coefficients can at most grow linearly in their indices, and chaotic systems exhibit this fastest linear growth. This linear growth is consistent with the chaos bound \cite{Maldacena:2015waa} and has attracted substantial studies in recent times \cite{Barbon:2019wsy,  Avdoshkin:2019trj, Dymarsky:2019elm, Rabinovici:2020ryf,  Cao:2020zls, Caputa:2021sib, Bhattacharya:2022gbz,  Bhattacharjee:2022lzy, Bhattacharya:2023zqt,  Liu:2022god, Hornedal:2022pkc, Fan:2022xaa, Bhattacharjee:2022ave, Alishahiha:2022anw, Avdoshkin:2022xuw, Camargo:2022rnt, Hornedal:2023xpa,  Erdmenger:2023wjg, Nizami:2023dkf, Patramanis:2023cwz, Camargo:2023eev,
Iizuka:2023pov, Bhattacharyya:2023dhp, Mohan:2023btr}, along with its cousins, circuit complexity and holographic complexity \cite{Chapman:2021jbh}.

On the other hand, open quantum systems are quantum systems that interact with their environment, which causes decoherence and dissipation. These effects are common in nature and have practical implications for quantum many-body systems. In particular, the study is important in the context of the black hole information problem in AdS/CFT correspondence, where the Hawking radiation is collected in a bath that is attached to an AdS black hole \cite{Almheiri:2020cfm}. In recent times, this has led to a surge of studies on the operator dynamics in open quantum systems, from quantum many-body systems \cite{Shibata:2018bir, Zhang:2022knu, Schuster:2022bot, Weinstein:2022yce} to quantum field theory and holography \cite{Loganayagam:2022zmq}. 

However, the operator evolution in an open system is drastically different compared to a closed system. The primary reason is the interaction with the environment, which makes the whole evolution non-unitary. This poses a challenge for studying operator growth in open systems, as the usual Lanczos algorithm does not work \cite{Bhattacharya:2022gbz}. The problem can be circumvented by applying more generic algorithms such as Arnoldi iteration \cite{Arnoldi1951ThePO} or the bi-Lanczos algorithm. A generic study of Krylov complexity was initiated in \cite{Bhattacharya:2022gbz, Bhattacharjee:2022lzy, Bhattacharya:2023zqt} using such algorithms. Especially, the authors of this paper initiated a study in the dissipative Sachdev-Ye-Kitaev (SYK) model \cite{Kulkarni:2021gtt} using Arnoldi iteration and motivated some universal aspects of the growth and saturation of Krylov complexity \cite{Bhattacharjee:2022lzy}. This study is relevant since this particular dissipative SYK model can be interpreted as two non-Hermitian SYK models coupled by a Keldysh wormhole \cite{Garcia-Garcia:2022adg}, suggesting that generic open quantum systems may have a holographic dual that involves a wormhole or a similar structure. However, a special form of dissipation is chosen, which is simple and thus not generic. Therefore, it is still an open question as to how robust the results are for other forms of dissipation.

In this paper, we partially answer this question by studying a large class of dissipators. By choosing a $p$-body Lindblad dissipator with Gaussian strength, in the large $N$ and the large $q$ limit, we analytically show that both the diagonal and off-diagonal coefficients of the Lindbladian matrix exhibit asymptotically linear growth, consistent with the observation made in spin chains \cite{Bhattacharya:2023zqt}. This is further supported by the results of the bi-Lanczos algorithm in the finite $N$ and finite $q$ SYK in the appropriate regime, modulo the finite-size effects. The resulting logarithmic timescale of dissipation and the saturation of Krylov complexity are found to be fairly general and independent of the choice of the form of the dissipation. We find that this growth and saturation are also reflected in the behavior of OTOC and operator size, which is supposed to construct a larger class of $\mathfrak{q}$-complexity measures \cite{Parker:2018yvk}. The saturation can also be interpreted as a result of continuous measurement by the environment itself. Finally, we also provide a generic notion of the pole structure of auto-correlation and the high-frequency behavior of the spectral function in the presence of dissipation. This leads us to motivate an operator growth hypothesis in generic open quantum systems.

Our paper is structured as follows. In section \ref{sec:system}, we introduce our system and the generic form of the dissipation. Section \ref{sec:biLanczos} introduces the machinery of the bi-Lanczos algorithm and the general version of the Krylov complexity in dissipative systems. In section \ref{analind}, we provide an analytical derivation of the linear growth of the diagonal coefficients of the Lindbladian matrix for any generic dissipation which we numerically confirm by implementing the bi-Lanczos algorithm in section \ref{numlind}. Based on the above results, section \ref{uni} motivates some universal aspects of Krylov complexity and its associated quantities. Finally, in section \ref{otocsec}, we derive the expression of OTOC for a $1$-body and general $p$-body fermionic initial operators. We conclude in section \ref{conclusion} with a brief summary and future outlook. Appendices consist of some further results and detailed derivation which we omit in the main text.

%The authors found a transition from the exponential to a power-law decay in the presence of dissipation, backed by computation with large $N$ but \emph{finite} $q$ dissipative SYK with a very specific choice of dissipation. However, in our study, we do not observe any power-law decay. The difference lies in choosing the auto-correlation function. While our choice is consistent with the asymptotic growth of Lanczos coefficients, the choice in  Ref.\,\cite{NSSrivatsa:2023qlh} involves a coarse-grained large $q$ summation that is technically valid in the first subleading order in $1/q$ expansion.  

\section{System and the environment: The Lindbladian} \label{sec:system}

The prototypical example of an open system consists of a system which is interacting with a dissipative environment. Our system under study is the Sachdev-Ye-Kitaev (SYK) model \cite{PhysRevLett.70.3339, Kittu}. This model has garnered much attention in recent times, especially being maximally chaotic \cite{Cotler:2016fpe, Garcia-Garcia:2017bkg}, and sharing the same Schwarzian action as Jackiw–Teitelboim (JT) gravity in low temperatures which elevates it as a toy model for holography. In addition, the generic model is known to be maximally chaotic, satisfying the Maldacena-Shenker-Stanford (MSS) bound \cite{Maldacena:2015waa}. From a condensed matter physics perspective, it provides detailed insights into non-Fermi liquids and strange metals \cite{Chowdhury:2021qpy}.

The $q$-body Sachdev-Ye-Kitaev (SYK) model consists of $N$ Majorana fermions $\psi_i$, satisfying the Clifford algebra $\{\psi_a, \psi_b\} = \delta_{ab}$, where $q$ fermions are interacting at a time. The Hamiltonian is given by \cite{PhysRevLett.70.3339, Kittu}
\begin{align}
    H = i^{q/2} \sum_{i_1 < \cdots < i_q} J_{i_1 \cdots i_q} \psi_{i_1} \cdots \psi_{i_q}\,, \label{sykh}
\end{align}
The random couplings $J_{i_1 \cdots i_q}$ are drawn from the Gaussian ensemble with the following mean and variance
\begin{align}
    \braket{J_{i_1 \cdots i_q}} = 0\,, ~~~~~~~ \braket{J^2_{i_1 \cdots i_q}} = \frac{(q-1)! J^2}{N^{q-1}} = 2^{1-q} \frac{(q-1)! \mathcal{J}^2}{q N^{q-1}}\,,
\end{align}
where $\mathcal{J}^2 = 2^{1-q} q J^2$. This notation is specifically useful in the large $q$ limit and $N\to\infty$ limit since the model is chaotic and becomes analytically tractable in this limit. As we have stated before, we will treat this as a system and examine various variations of this model in the open system setting in the following subsections.

We consider the system to be connected to an environment governed by Markovian dynamics. This regime is defined when the system density matrix $\rho(t+dt) = \rho(t) + O(dt)$ is solely determined by the system density matrix at time $t$ i.e., $\rho(t)$. More broadly, we consider the dissipative mechanism where the information leaks out from the system to the environment such that it never returns to the system at a later time.\footnote{In the generic case, the environment can also transfer some information to the system which results in a complicated non-Markovian evolution.} In other words, our observed time scale of the system dynamics $t_{\mathrm{sys}}$ is long compared to the timescale $\delta t_{E}$ that the environment retains the memory of the information that has been leaked out from the system i.e., $t_{\mathrm{sys}} \gg \delta t_{E}$. Under the Born-Markovian approximation, and weak coupling regime, the evolution of the density matrix and any operator can be treated in the realm of Lindbladian formalism \cite{Lindblad1976, Gorini}.
The density matrix of the system evolves by the master equation
\begin{align}
	\dot{\rho} =-i[H,\rho]+\sum_k \big[L_k \rho L_k^{\dagger}-\frac{1}{2}\{L_k^{\dagger} L_k,\rho\} \big]\,, \label{lstate}
\end{align}
where $H$ is the system Hamiltonian. The operators $L_k$ are referred to as Lindblad jump operators and they capture the information of the interaction between the system and the environment. In particular, they are made of system operators only and completely lack detailed information about the environment. An arbitrary initial operator $\mathcal{O}_0$ at $t=0$ evolves as
\begin{align}
    \mathcal{O}(t) = e^{i \mathcal{L}_o^{\dagger} t}\, \mathcal{O}_0\,. \label{adO}
\end{align}
Here $\mathcal{L}_o^{\dagger}$ is known as the (adjoint) Lindbladian for the operator\footnote{The dynamics of the density matrix is governed by the Lindbladian $\mathcal{L}_o$. In this paper, we continue calling $\mathcal{L}_o^{\dagger}$ as Lindbladian unless specified.} and acts as
\begin{align}
    \mathcal{L}_o^{\dagger} \mathcal{O} = [H, \mathcal{O}] - i \sum_k \left[\pm L_k^{\dagger} \mathcal{O} L_k - \frac{1}{2} \left\{L_k^{\dagger} L_k, \mathcal{O}\right\} \right]\,. \label{lindfull}
\end{align}
Here the ``$-$'' sign should be considered in case both $L_k$ and $\mathcal{O}$ are fermionic \cite{Liu:2022god}. One usually express it in a vectorized form after the Choi-Jami\l kowski isomorphism \cite{CHOI1975285, JAMIOLKOWSKI1972275} with the following replacement \cite{horn_johnson_1991, am2015three}
\begin{align}
    A \,\mathcal{O}\, B \rightarrow (B^T \otimes A) (\mathrm{vec} \,\mathcal{O})\,,
\end{align}
where $A$ and $B$ are any arbitrary operators and $\mathrm{vec}\, \mathcal{O}$ is the vectorization of the operator $\mathcal{O}$. This gives the Lindbladian superoperator
\begin{align}
    \mathcal{L}_o^{\dagger} \equiv (I \otimes H - H^{T} \otimes I) - i \sum_k \left[\pm L_k^T \otimes L_{k}^{\dagger} - \frac{1}{2} \left(I \otimes L_{k}^{\dagger} L_k + L_k^T L_k^{*} \otimes I\right) \right]\,, \label{lindsup}
\end{align}
%\begin{align}
%    \mathcal{L}_o = (I \otimes H - H^{T} \otimes I) - i \sum_m \left[\pm L_m^T \otimes L_{m}^{\dagger} + \frac{1}{2} \left(I \otimes L_{m}^{\dagger} L_m + L_m^T L_m^{*} \otimes I\right) \right]\,,
%\end{align}
The notation ``$\equiv$'' indicates that the Lindbladian is represented in a matrix form in the doubled-Hilbert space. In this paper, our system is the SYK Hamiltonian \eqref{sykh} and we take 
the following two classes of jump operators:

\emph{Class 1: Linear dissipator:} We consider an open system version of SYK with the following jump operators~\cite{Kulkarni:2021gtt}:
\begin{align}
    L_{i} = \sqrt{\lambda} \,\psi_{i}\,, ~~~~~~ i = 1, 2, \cdots, N\,. \label{jumpop}
\end{align}
with $\lambda \geq 0$ being the coupling strength between the system and the environment. We often call the full system a dissipative SYK with linear jump operators. This is the simplest version of dissipative SYK, where each of the fermions dissipates at an equal rate. One can solve this model analytically in the large-$q$ limit. This model is particularly useful in being realized as a connected Keldysh wormhole \cite{Garcia-Garcia:2022adg}. A detailed study of Krylov complexity in this setup was conducted in \cite{Bhattacharjee:2022lzy}. This analysis can be extended to a dissipation strength of the type $V_{i}$ (instead of $\sqrt{\lambda}$) where $V_{i}$ are Gaussian random complex numbers with zero mean and finite variance. The results are the same under disorder averaging, up to an appropriate identification of the respective parameters.

\emph{Class 2: Non-linear dissipator:} For this class, we take the $p$-body jump operators of the following form \cite{Kawabata:2022cpr}
\begin{align}
    L_a = \sum_{1 \leq i_1 < \cdots < i_p \leq N} V_{i_1 i_2 \ldots i_p}^a \,\psi_{i_1} \psi_{i_2} \cdots \psi_{i_p}\,, ~~~~ a =1, 2, \cdots, M\,, \label{jpp0}
\end{align}
with the following distribution of $V_{i_1 i_2 \ldots i_p}$:
\begin{align}
    \langle V_{i_1 i_2 \ldots i_p}^a \rangle = 0\,, ~~~~~ \langle |V_{i_1 i_2 \ldots i_p}^a|^2 \rangle = \frac{p!}{N^p}  V^2\,,~~~~ \forall i_1, \cdots, i_p, a\,, \label{jp}
\end{align}
with $V \geq 0$. In other words, the jump operators are $p$-local and mimic the SYK-like structure. Together with the Hamiltonian \eqref{sykh}, they dictate the full non-unitary dynamics governed by the Lindbladian
\eqref{lindsup}. In particular, the parameter $J$ represents the unitary dynamics while the parameter $V$ breaks it. The $p = 1$ case without the random average is known as the linear dissipator (\emph{class 1}). The $p = 2$ case is known as the quadratic dissipator model, previously introduced in \cite{Kulkarni:2021gtt, Sa:2021tdr}. In the following sections, our interests will be the generic $p$-body dissipator with a possible emphasis on linear ($p=1$) and quadratic ($p=2$) dissipator cases particularly.\footnote{The jump operators chosen here encompass a large class of non-trivial Markovian dissipative operators, which are random. As described in \cite{Kawabata:2022cpr}, they are quite generic. However, they are not the most general form, especially when one can consider $p$-body operators without the sum and/or randomness. We believe that randomness is crucial for our purpose, and hence we only focus on the class $2$ operators with random averaging with the sum.}

Before jumping on to the numerical machinery of the bi-Lanczos algorithm, we briefly discuss a different perspective of the Lindblad (and Lindblad-like) equation \cite{Zhou:2022qhe}. Suppose we prepare the system at time $t$ with a density matrix given by $\rho(t)$ and evolve unitarity till time $t+\delta t$. The state of the system is $\rho(t+\delta t)$, which can be written as
\begin{align}
    \rho(t+\delta t) = \rho(t) - i [H, \rho (t)] \delta t + O(\delta t^2)\,. \label{st1}
\end{align}
This equation purely comes from the unitary dynamics of the system, namely the Schrodinger equation $d\rho (t)/dt = - i [H, \rho (t)]$. However, if we measure the system with a probability $P(t+\delta t)$ at time $t+\delta t$, then the state after the measurement is given by
\begin{align}
    \rho^M(t+\delta t) = [1 - P(t+\delta t)] \rho(t+\delta t) + P(t+\delta t) \sum_k L_k \rho(t+\delta t)  L_k^{\dagger}\,, \label{st2}
\end{align}
where $L_k$'s are the same quantum jump operators as introduced in \eqref{lindfull}. They act as projector operators satisfying the completeness relation $\sum_k L_k^{\dagger} L_k = I$.\footnote{This condition is very special and can be relaxed for the ``weak measurements'' \cite{PhysRevLett.60.1351}.} We further expand the probability as
\begin{align}
     P(t+\delta t) = P(t) + \eta (t) \delta t + O(\delta t^2)\,, \label{st3}
\end{align}
where $\eta (t) = \delta P(t)/\delta t$ is the change of measurement probability in unit time and we refer to it as the measurement rate. Plugging \eqref{st1} and \eqref{st3} into the expansion of \eqref{st2}, and neglecting $O(\delta t^2)$ terms, we obtain
\begin{align}
    \frac{\partial \rho^M (t)}{\partial t} = - i [H, \rho(t)] + \eta (t) \sum_k \left[L_k \rho(t) L_k^{\dagger} - \frac{1}{2} \{L_k^{\dagger} L_k, \rho(t)\}\right]\,, \label{st4}
\end{align}
which has a surprisingly similar form to the Lindblad master equation for the density matrix. The anti-commutator part is trivial here due to the completeness relation. The above expression \eqref{st4} bears a physical significance. In particular, we can directly associate the dissipation strength in \eqref{jumpop} or \eqref{jp} as the measurement rate $\eta (t)$ by the environment itself. Depending on the strength of the dissipation, the system can be either in a fully scrambled phase or a purely dissipative phase, opening the possibility of studying the so-called ``environment-induced phase transition'' \cite{Zhang:2022knu}.

\section{Bi-Lanczos algorithm for open systems} \label{sec:biLanczos}

As we briefly discussed in the introduction section, the bi-Lanczos algorithm is a numerical method that can transform a Lindbladian matrix into a tri-diagonal form, which can be used to compute the Lanczos coefficients and the Krylov complexity of an open quantum system. The bi-Lanczos algorithm was first applied to the study of Krylov complexity in \cite{Bhattacharya:2023zqt}, where some properties of the coefficients were studied in spin chains. In this section, we will review the bi-Lanczos algorithm and present some more properties of the coefficients, such as their asymptotic behavior. We will then apply the bi-Lanczos algorithm to the dissipative SYK model in later sections, and compare our results with analytical counterparts.

\subsection{Vectorization and bi-orthonormal vectors}\label{biLsec}
The central idea of the bi-Lanczos algorithm is to construct two sets of bi-orthonormal vectors $|p_n\rrangle$ and $|q_n\rrangle$ satisfying the following bi-orthonormal condition
\begin{equation}\label{biortho}
    \llangle q_m|p_n \rrangle =\delta_{mn}\,.
\end{equation}
We use the notation of \cite{Bhattacharya:2023zqt} to denote the bi-Lanczos vectors with ``double braces''. They are obtained by vectorizing the initial operator. These two bi-orthonormal vectors evolve differently under the Lindblad evolution. In the absence of dissipation, the Lindbladian reduces to the Liouvillian, which is Hermitian and can be recast into a purely tridiagonal form with vanishing diagonal coefficients. The two spaces become conjugate to each other and thus become individually orthonormal \cite{Parker:2018yvk}. Now, we outline the steps of the bi-Lanczos algorithm \cite{Bhattacharya:2023zqt, gaaf2017infinite}:\footnote{We have made our notation slightly different and more compact than \cite{Bhattacharya:2023zqt}.}

\begin{enumerate}
    \item \textbf{Initialization.}
    
    Let $|p_0\rrangle = |q_0\rrangle = 0$ and $b_0 = c_0 = 0$. Also, let  $|p_1\rrangle = |q_1\rrangle \equiv |\mathcal{O}_0)$, where $\mathcal{O}_0$ is the initial vector.
    \item \textbf{Lindbladian action and bi-Lanczos coefficients.}

    For $j = 1, 2, \ldots$, we perform the following iterations:
    \begin{enumerate}
        \item Compute:  $|r_j\rrangle =\mathcal{L}_{o}^\dagger  |p_j\rrangle$ and $|s_j\rrangle =\mathcal{L}_{o} |q_j\rrangle$.
        
        \item Redefine the vectors:
        
        $|r_j\rrangle := |r_j\rrangle - b_{j-1} |p_{j-1}\rrangle$ and $|s_j\rrangle := |s_j\rrangle - c_{j-1}^{*} |q_{j-1}\rrangle$.
        
        \item  Evaluate the inner product: $a_j = \llangle q_j|r_j \rrangle$.

        \item Again, redefine the vectors:
        
        $|r_j\rrangle := |r_j\rrangle - a_j |p_{j}\rrangle$ and $|s_j\rrangle := |s_j\rrangle - a_j^{*} |q_{j}\rrangle$.

        \item Evaluate the inner product: $\omega_{j}=\llangle r_{j}|s_{j} \rrangle$.

        \item Evaluate the norm:  $c_{j}= \sqrt{|\omega_{j}|}$ and $b_{j} = \omega^*_{j}/c_{j}$.

        \item If $c_{j+1}\neq0$, then define the vectors:
        \begin{align}
            |p_{j+1}\rrangle = \frac{|r_{j}\rrangle}{c_{j}}~~~~ \mathrm{and} ~~~~ |q_{j+1}\rrangle =\frac{|s_{j}\rrangle}{b^*_{j}}\,.
        \end{align}

        \item Check the convergence and perform the full orthogonalization (FO) procedure, if required.
        %\footnote{Sometimes, it is instructive to perform the FO twice to obtain better numerical accuracy \cite{Rabinovici:2020ryf}.} if required:
        %\begin{align*}
        %    |p_{j+1}\rrangle = |p_{j+1}\rrangle - \sum_{i=1}^{j} \llangle q_i|p_{j+1} \rrangle\, |p_{j+1} \rrangle\,, ~~~~~~
         %   |q_{j+1}\rrangle = |q_{j+1}\rrangle - \sum_{i=1}^{j} \llangle p_i|q_{j+1} \rrangle \, |q_{j+1} \rrangle\,.
        %\end{align*}
\end{enumerate}
        \item Stop, if $c_k=0$ for some $k$.
\end{enumerate}
The algorithm generates three sets of coefficients $\{a_j\}, \{b_j\}$ and $\{c_j\}$, and two sets of bi-orthogonal vectors $\{|p_j\rrangle\}$ and $\{|q_j\rrangle\}$. The full action of this bi-Lanczos basis can be expressed in the following form, which are two sets of three-term recurrences
\begin{align}
    c_{j} |p_{j+1}\rrangle &=\mathcal{L}_{o}^\dagger |p_j\rrangle -a_j |p_j \rrangle  -b_{j-1} |p_{j-1}\rrangle\,, \label{bilanczosbasic} \\
    b^*_{j} |q_{j+1}\rrangle &= \mathcal{L}_{o} |q_j\rrangle - a^*_j |q_j\rrangle -c^*_{j-1} |q_{j-1}\rrangle\,, \label{bilanczosbasic2}
\end{align}
where $*$ denotes the complex conjugate. In other words, we have generated two sets of Krylov spaces, one acts by $\mathcal{L}_o$ and the other one by $\mathcal{L}^{\dagger}_o$: 
\begin{align} \mathrm{Kry}^j(\mathcal{L}_{o}^\dagger,\,|p_1\rrangle)&= \{|p_1\rrangle, \mathcal{L}_{o}^\dagger \, |p_1\rrangle, (\mathcal{L}_{o}^\dagger)^2 \,|p_1\rrangle, \ldots\}\,,   \\
  \mathrm{Kry}^j(\mathcal{L}_{o},\,|q_1\rrangle)&= \{|q_1\rrangle, \mathcal{L}_{o} \, |q_1\rrangle, \mathcal{L}_{o}^2 \, |q_1 \rrangle, \ldots \}\,.
\end{align}
From the recurrence \eqref{bilanczosbasic}, it is evident that the procedure of the bi-Lanczos algorithm recasts the Lindbladian into the following  tridiagonal form
\begin{align}
    \mathcal{L}_{o}^{\dagger} \equiv \begin{pmatrix} a_{1}&b_{1}&&&&0\\c_{1}&  a_{2}& b_{2}&&&\\&c_{2}&\ddots&\ddots &&\\&&\ddots &a_{m} &b_{m}&\\&&&c_{m}&\ddots&
    \ddots\\0&&&&\ddots&\ddots\\\end{pmatrix}\,.
\end{align}
In other words, we have elements in the diagonal and the primary off-diagonal terms only. This is a distinctive feature from the Arnoldi iteration \cite{Bhattacharya:2022gbz, Bhattacharjee:2022lzy} which considered only a single set of orthonormal vectors and renders the Lindbladian into an upper-Hessenberg form. However, the methodology to generate the orthonormal vectors is different in the Arnoldi iteration where the individual space is orthonormal. Also, the numerical stability significantly differs in both cases. While the computational cost (time complexity) is higher in the Arnoldi iteration compared to the bi-Lanczos algorithm, the latter might suffer a breakdown
i.e., the loss of orthogonality. Such breakdown never occurs in Arnoldi iteration \cite{sogabe2023krylov}.

As we will see in the upcoming sections, our numerical algorithm implies $b_n = c_n$. The Lindbladian matrix can be written in tridiagonal form in Krylov (bi-Lanczos) basis in the following form
\begin{align}
    \mathcal{L}_{o}^{\dagger} \equiv \begin{pmatrix} a_{1}&b_{1}&&&&0\\b_{1}&  a_{2}& b_{2}&&&\\&b_{2}&\ddots&\ddots &&\\&&\ddots &a_{m} &b_{m}&\\&&&b_{m}&\ddots&
    \ddots\\0&&&&\ddots&\ddots\\\end{pmatrix}\,. \label{ld}
\end{align}
We find that the diagonal coefficients $a_n$ are purely imaginary $a_n = i |a_n|$ and the off-diagonal elements $b_n$ are purely real. This implies that the Lindbladian $\mathcal{L}_{o}^{\dagger}$ is neither Hermitian ($\mathcal{L}_{o}^{\dagger} \neq (\mathcal{L}_{o}^{\dagger})^{\dagger}$) nor anti-Hermitian ($\mathcal{L}_{o}^{\dagger} \neq -(\mathcal{L}_{o}^{\dagger})^{\dagger}$). It is generically non-Hermitian. For more details of the properties of these coefficients, see \cite{Bhattacharya:2023zqt} and subsection \ref{biprop}. 

\begin{figure}[t]
   \centering
\includegraphics[width=0.95\textwidth]{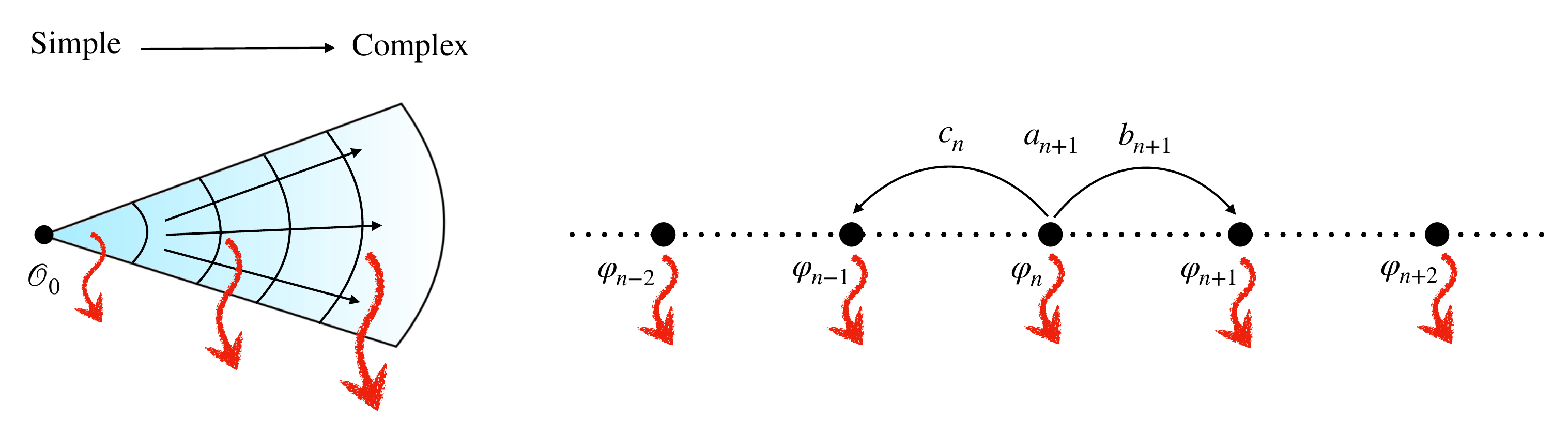}
\caption{The operator growth in dissipative systems (left) can be mapped to a model of the non-Hermitian Krylov chain (right). The hopping amplitudes from $n$-th side to $(n+1)$-th and $(n-1)$-th sites are $b_{n+1}$ and $c_n$ respectively, while $a_{n+1}$ gives the amplitude of staying at site $n$. Here $\mathcal{O}_0$ indicates the initial operator and the red arrows indicate (increasing length indicates stronger dissipation) the dissipation which affects all sites.} \label{fig:nonHchain}
\end{figure}

\subsection{Operator growth and Krylov complexity}

Since the operator evolves with $\mathcal{L}_o^{\dagger}$, we expand the time-evolved operator in the bi-Lanczos basis in the following form
\begin{align}
    |\mathcal{O}(t)) = \sum_n i^n \varphi_{n-1} (t) \,|p_n\rrangle\,,
    %~~~~~~~~  (\mathcal{O}(t)| = \sum_n (-i)^n \psi_{n-1} (t)  \llangle q_n|\,.
\end{align}
where $\varphi_n$ are Krylov basis wavefunctions. We slightly changed the notation from \cite{Parker:2018yvk} to keep the auto-correlation function as $\varphi_0$ since our initial vector starts with $|p_1\rrangle$ (and $|q_1\rrangle$) instead of $|p_0\rrangle$ (and $|q_0\rrangle$). Heisenberg equation of motion $d |\mathcal{O}(t))/d t = i \mathcal{L}_o^{\dagger} |\mathcal{O}(t))$ becomes
\begin{align}
    \partial_t \varphi_{n-1} = c_{n-1} \varphi_{n-2} + i a_n \varphi_{n-1} - b_n \varphi_n \,, ~~~~ n \geq 1\,. \label{nonHtight}
\end{align}
where $\varphi_n \equiv \varphi_n (t)$ for brevity and $\varphi_0$ is the auto-correlation function. For the Lindbladian evolution, it is defined as
\begin{align}
    \varphi_0(t) \equiv C(\{\mu\}, t) = \frac{1}{2^N} \mathrm{Tr}(\mathcal{O}(t) \mathcal{O}_0)\,,
\end{align}
where $\mathcal{O}(t)$ is given by Eq.\,\eqref{adO}, $\{\mu\}$ is the set of the dissipative parameters, and $N$ is the number of degrees of freedom in the system. Eq.\,\eqref{nonHtight} can be interpreted as a non-Hermitian tight-binding model \cite{Liu:2022god}, which we refer to as the non-Hermitian Krylov chain (see Fig.\,\ref{fig:nonHchain}). The particle hops from $n$-th site to $(n+1)$-th site with hopping amplitude $b_{n+1}$ and to $(n-1)$-th site with a different amplitude $c_n$, while the amplitude of staying at that particular site is $a_{n+1}$. As we will examine later, in all the examples of various versions of dissipative SYK, we numerically find $b_n = c_n$ for all $n$. Moreover we find all $a_n$ are imaginary, i.e., $a_n = i |a_n|$. Hence the Eq.\,\eqref{nonHtight} simplifies to\footnote{In order to match \eqref{nonHtight} with the relevant equation in \cite{Bhattacharya:2023zqt}, we first rescale $a_n \rightarrow a_{n-1}$ since our first element is $a_1$ whereas \cite{Bhattacharya:2023zqt} uses the first element as $a_0$. Then we shift $n \rightarrow n+1$ and start with $n =0$. For closed systems, $a_n = 0$ which further reduces to the Hermitian Krylov chain \cite{Parker:2018yvk}.}
\begin{align}
    \partial_t \varphi_{n-1} = b_{n-1} \varphi_{n-2} -  |a_n| \varphi_{n-1} - b_n \varphi_n \,, ~~~~ n \geq 1\,.
\end{align}
The Krylov complexity is thus defined as the average position of a particle in the non-Hermitian Krylov chain given by
\begin{align}
    K(t) = \frac{1}{\mathcal{Z}}  \sum_n n |\varphi_n (t)|^2 = \frac{\sum_n n |\varphi_n (t)|^2}{ \sum_n  |\varphi_n (t) |^2}  \,,
\end{align}
where $\mathcal{Z} =  \sum_n  |\varphi_n (t) |^2$ is the normalization. The probability $\sum_n  |\varphi_n (t) |^2 < 1$ is not conserved due to the unitarity breaking, thus a division is required since the rescaled amplitude $\varphi (t)/\sqrt{\mathcal{Z}}$ conserves the probability i.e., $\sum_n  |\varphi_n (t)/\sqrt{\mathcal{Z}} |^2 = 1$.

\subsection{Properties of bi-Lanczos coefficients} \label{biprop}

The bi-Lanczos algorithm generates three sets of coefficients. In general, all the coefficients can be complex numbers. However, as we see the structure and physical properties of a Lindbladian matrix heavily restrict the properties of these coefficients. The density matrix evolution is governed by $\mathcal{L}_o$, while the operator evolution is governed by the adjoint $\mathcal{L}_o^{\dagger}$ \cite{Shibata:2018bir}. This will be reflected transparently in the structure of Lindbladian in the bi-Lanczos basis. Specifically, $\mathcal{L}_o^{\dagger}$ has positive and purely imaginary diagonal coefficients, while the diagonal coefficients of $\mathcal{L}_o$ are negative and purely imaginary. In either case, the off-diagonal coefficients are purely real. Together, this makes $\mathcal{L}_o^{\dagger}$ or $\mathcal{L}_o$ non-Hermitian. The structure of these elements must be consistent such that the eigenvalues of $- i \mathcal{L}_o^{\dagger}$ or $i \mathcal{L}_o$ have to be either purely real positive or complex conjugate in pairs with the real part being positive.\footnote{To clarify more, for an even-dimensional matrix, all the eigenvalues of $- i \mathcal{L}_o^{\dagger}$ or $i \mathcal{L}_o$ have to be complex conjugate in pairs i.e., of the form $\alpha \pm i \beta$, with $\alpha > 0$. The purely real positive eigenvalue appears only for an odd-dimensional matrix.} In all the examples we study, these conditions will be fulfilled. However, before delving into the examples, we state a proposition\footnote{We also wonder if such inequality can be leveraged to understand the Liouvillian gap and relaxation time in Markovian open quantum systems \cite{Mori:2022qbc, Shibata:2018bir}. We leave it to future work.} concerning the elements of the bi-Lanczos coefficients.\\
\newline
\textbf{Proposition 1.} \emph{The imaginary part of any eigenvalue $\lambda_{\mathcal{L}}$ of $\mathcal{L}_{o}^\dagger$ satisfies
\begin{align}
    \underset{1 \leq n \leq \mathcal{K}}{\mathrm{min}} \,\mathrm{Im}(a_n) \,\leq \,\mathrm{Im}(\lambda_{\mathcal{L}})\, \leq\, \underset{1 \leq n \leq \mathcal{K}}{\mathrm{max}} \,\mathrm{Im}(a_n)\,. \label{theorem}
\end{align}
where $\mathcal{K}$ is the Krylov dimension. For a closed system, $\lambda_{\mathcal{L}}$ of $\mathcal{L}_{o}^\dagger$ is Hermitian, and all $a_n$ vanishes, thus \eqref{theorem} holds trivially with equality.}\\
\newline
\emph{Proof:} Following \cite{theorem}, we consider a diagonal matrix $D$, such that we transform the matrix \eqref{ld} into the following form
\begin{align}
    D^{-1}[\mathcal{L}_{o}^{\dagger}]D = \begin{pmatrix} a_{1}&\sqrt{b_{1} c_1} &&&&0\\ \sqrt{b_1 c_{1}}&  a_{2}& \sqrt{b_{2} c_2} &&&\\& \sqrt{b_2 c_{2}}&\ddots&\ddots &&\\&&\ddots &a_{m} &\sqrt{b_{m} c_m} &\\&&&\sqrt{b_m c_{m}}&\ddots&
    \ddots\\0&&&&\ddots&\ddots\\\end{pmatrix}\,. \label{ld01}
\end{align}
Assume $\boldsymbol{\xi} = (\xi_1 \cdots \xi_m \cdots)^T$ be as a unit eigenvector of $D^{-1}[\mathcal{L}_{o}^{\dagger}]D$ associated to an eigenvalue $\lambda_{\mathcal{L}}$. Then the eigenvalue reads
\begin{align}
    \lambda_{\mathcal{L}} = \boldsymbol{\xi}^{\dagger} \big(D^{-1}\,[\mathcal{L}_{o}^{\dagger}]D \big) \,\boldsymbol{\xi} = \sum_n a_n |\xi_n|^2 + \sum_n \sqrt{b_n c_n} (\xi^{*}_n \xi_{n+1} + \xi_n \xi^{*}_{n+1})\,,
\end{align}
where the sum runs over the corresponding elements (up to the Krylov dimension). Thus we obtain
\begin{align}
    \mathrm{Im} (\lambda_{\mathcal{L}}) = \mathrm{Im} \left(\sum_n a_n |\xi_n|^2\right) =  \sum_n \mathrm{Im} (a_n) \, |\xi_n|^2 \,.
\end{align}
Hence, \eqref{theorem} follows. 

The form of the Lindbladian \eqref{ld01} suggests an alternate form of Lanczos coefficients $d_n := \sqrt{b_n c_n}$, which was recently advocated in \cite{NSSrivatsa:2023qlh}.

\section{Lindbladian SYK: analytical treatment} \label{analind}

In this section, we provide a simple analytical treatment to show the linear growth of diagonal coefficients. We split the Lindbladian \eqref{lindfull} into two parts $\mathcal{L}_o^{\dagger} \,\mathcal{O}= \mathcal{L}_{H}^{\dagger} \,\mathcal{O} + \mathcal{L}_{D}^{\dagger}\, \mathcal{O}$, namely
\begin{align}
    \mathcal{L}_{H}^{\dagger} \,\mathcal{O} = [H,\mathcal{O}]\,, ~~~~~~
    \mathcal{L}_{D}^{\dagger} \,\mathcal{O} =  - i \sum_{k = 1}^{M}[\pm L^\dagger_{k}\mathcal{O}L_{k} - \frac{1}{2}\{L^\dagger_{k}L_{k},\mathcal{O}\}]\,. \label{splind}
\end{align}
Here we need to choose the ``-'' sign if both the jump operators and the initial operators are fermionic. Our derivation is based on a property known as the ``operator size concentration'' \cite{Bhattacharjee:2022lzy}. It states that the eigenoperators of the dissipative part $\mathcal{L}_{D}^{\dagger}$ (ignoring $o(1/q)$ corrections) are given by 
\begin{align}
    \mathcal{O}_{n} = \sum_{i_{1}<i_{2}<\dots<i_{s}}c_{i_{1} i_{2} \cdots i_{s}}\psi_{i_{1}}\psi_{i_{2}}\cdots \psi_{i_{s}} + o(1/q)\label{Onterm}
\end{align}
where $c_{i_{1} i_{2} \cdots i_{s}}$ are some coefficients, and $s = n(q-2)+1$.  This property only holds in the large $q$ and large $N$ limit, hence our analytical derivation only holds in this limit.
\subsection{For linear dissipator}
We will present our first example of this model in the open-system setting. This is done by introducing the linear jump operators of the form~\cite{Kulkarni:2021gtt}:
\begin{align}
    L_{i} = \sqrt{\lambda} \,\psi_{i}\,, ~~~~~~ i = 1, 2, \cdots, N\,. \label{linj}
\end{align}
with $\lambda \geq 0$. These are the jump operators of \emph{class 1}, as we have discussed earlier. For the case of the fermionic jump operator, it was shown in \cite{Bhattacharjee:2022lzy} that the dissipative part of the Lindbladian acts linearly i.e., 
\begin{equation}
    \mathcal{L}_D^{\dagger} \mathcal{O}_n = i \lambda  s \, \mathcal{O}_n = i \tilde{\lambda} n \,\mathcal{O}_n \label{eq:LDMajorana} \,.
\end{equation}
where $\tilde{\lambda} = \lambda q$. This immediately gives $a_n \sim i \tilde{\lambda} n$. Moreover, the advantage of these jump operators is that they allow us to perform an exact analytical calculation in the large-$q$ limit. In particular, one can solve the Schwinger-Dyson equation and obtain the following two-point function~\cite{Kulkarni:2021gtt}:
\begin{align}
  &  C(\tilde{\lambda},t) =  1 + \frac1q g(t) + o(1/q^2) \,, \label{eq:Ct} \\
  & g(t) =  \log \left[\frac{\alpha^2}{\mathcal{J}^2\cosh^2(\alpha t + \aleph)}\right] \,,\, ~~t > 0 \,, \label{ect1}
\end{align}
where $\tilde{\lambda} = \lambda q$ is a redefined coupling in the large-$q$ limit. The parameters $\alpha$ and $\aleph$ are related to the couplings as 
\begin{align}
     \alpha = \sqrt{( \tilde{\lambda}/ 2 )^2 +  \mathcal{J}^2}\,, ~~~~ \aleph = \arcsinh (\tilde{\lambda}/ (2 \mathcal{J}))\,. \label{pardef}
\end{align}
The closed-system result recovers in the limit $\lambda = 0$ and thereby obtaining $g(t) = 2 \ln \sech (\mathcal{J} t)$, which is a known result~\cite{Maldacena:2016hyu}.  We can also compute the Lanczos coefficients using the moment method \cite{viswanath1994recursion}. They are given by \cite{Bhattacharjee:2022lzy}
\begin{align}
&a_n = i \tilde{\lambda} n + o(1/q) \,,\,~~ \tilde{\lambda} := \lambda q \,,\, \label{eq:an-main} \\ &b_n =  \begin{cases} \mathcal{J}\sqrt{2/q}  \,      & \,n = 1\,,\\
    \mathcal{J}\sqrt{n(n-1)} + o(1/q) \,   & \,n > 1\,. \label{eq:bn-main}
  \end{cases} 
 \end{align}
Not only does the expression of $b_n$ match exactly with the closed-system result \cite{Parker:2018yvk}, but also the expression of $a_n$ matches what we obtained from the ``operator size concentration'' property, with an $o(1/q)$ correction which vanishes in the large $q$ limit. This gives the first hint that the off-diagonal elements of the Lindbladian matrix \eqref{ld} might not depend on the dissipation which is entirely reflected in the diagonal coefficients. However, the linear dissipation \eqref{linj} is very special and it is unclear if the above conclusion is generic for any arbitrary dissipation. Thus, we need to choose a more generic dissipation of \emph{class 2} to justify (or falsify) our conclusion.

\subsection{For random quadratic dissipator}
Next, we consider the jump operators which are random and quadratic. This belongs to the \emph{class 2} non-linear dissipator with $p=2$. In particular, we choose
\begin{align}
    L^a = \sum_{1 \leq i < j \leq N} V_{ij}^a \,\psi_{i} \psi_{j}\,, ~~~~ a =1, 2, \cdots, M\,,
\end{align}
with the following distribution of $V_{ij}$ drawn of random Gaussian ensemble
\begin{align}
    \langle V_{i j}^a \rangle = 0\,, ~~~~~ \langle |V_{ij}^a|^2 \rangle = \frac{2V^2}{N^2} \,~~~ \forall i,j,a\,. \label{vn}
\end{align}
In principle, $V$ can be arbitrary but for our computation, we focus on the weak-dissipation regime, which implies $J \gg V$.\footnote{Since we are working in finite $N$ and finite $q$, we take our system disorder parameter as $J$ instead of $\mathcal{J}$. The latter is important in the large $q$ limit.} This choice is motivated by the fact that we are considering the system dynamics that is Markovian. 

Similar to the linear case, we divide the Lindbladian into a Hermitian and a dissipative part. We choose the ``+'' sign in \eqref{splind} since the jump operators are bosonic. Since our primary concern is the dissipative part $\mathcal{L}_{D}^{\dagger}$, we consider the string $\psi_{i_{1}}\psi_{i_{2}}\dots \psi_{i_{s}}$ with $i_{1}<i_{2}<\dots<i_{s}$ and attempt to divine its' action on. In fact, the action of the dissipative part of the Lindbladian to a string of length $s$ results in the following proposition:\\
\newline
\textbf{Proposition 2.} \emph{Under ensemble averaging, the action of the dissipative part of the Lindbladian to a string of length $s = n(q-2)+1$ results in the following expression:
\begin{align}
    \mathcal{L}_{D}^{\dagger} \, \mathcal{O}_n = i \zeta q R V^2 n \, \mathcal{O}_n\,, \label{prop32}
\end{align}
with $R = M/N$. Here $\zeta \sim o(1)$ number and $V$ is given by the ensemble average \eqref{vn}}.

The proof is given in the  Appendix \ref{randquadproof}. The asymptotic linear growth can be deducted
\begin{align}
    a_n \sim i R V^2 \, n\,. \label{ac1}
\end{align}
In the large $q$ and large $N$, limit $R = M/N$ becomes the relevant quantity with $M$ being the number of jump operators.

%\subsection{For random $p$-body dissipator}
We can generalize the above result as a generic $p$-body dissipator of the form \eqref{jpp0}-\eqref{jp}. However, the computation is tedious and put in the Appendix \ref{randpbodyproof}. We obtain
\begin{align}
    \mathcal{L}_{D}^{\dagger} \, \mathcal{O}_n = i \frac{p s}{2^{p-1}}R V^2  \, \mathcal{O}_n\,, \label{prop4}
\end{align}
which is strictly valid in the large $N$ and large $q$ limit. It is easy to see that this reduces to the leading order contribution \eqref{prop32} for $p  = 2$. It is straightforward to conclude
\begin{align}
    a_n \sim i R V^2 \,n\,,
\end{align}
i.e., the asymptotic growth of the diagonal coefficients is linear and similar to \eqref{ac1}.

\section{Bi-Lanczos algorithm in Lindbladian SYK} \label{numlind}

To justify the above analytical results, we resort to the numerical bi-Lanczos algorithm to transform the Lindbladian into a pure tridiagonal form as given by equation \eqref{ld}. We separately apply this algorithm for linear, quadratic, and cubic dissipators.

\begin{figure}[t]
   \centering
\begin{subfigure}[b]{0.46\textwidth}
\centering
\includegraphics[width=\textwidth]{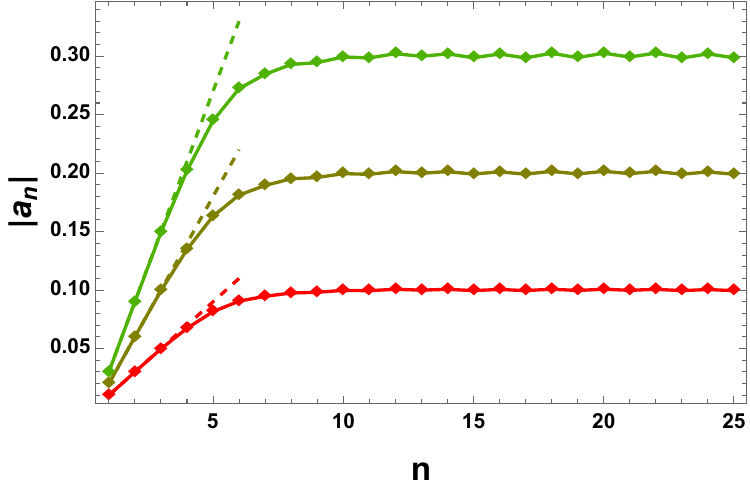}
\caption{Behavior of $|a_n|$.}
\end{subfigure}
\hfill
\begin{subfigure}[b]{0.46\textwidth}
\centering
\includegraphics[width=\textwidth]{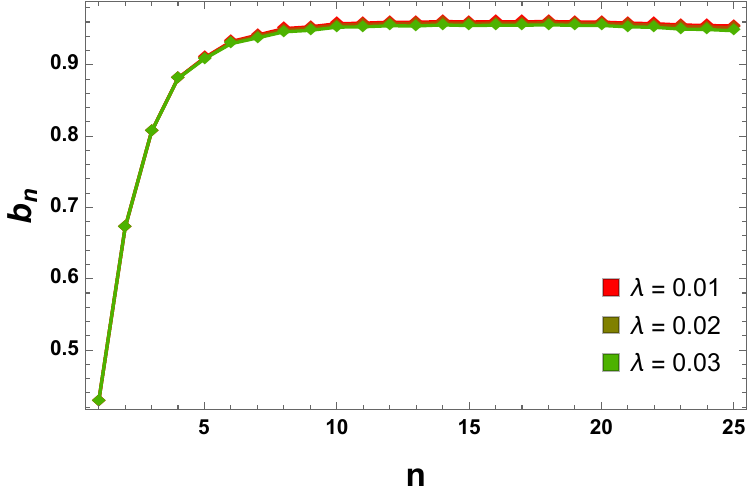}
\caption{Behavior of $b_n$.}
\end{subfigure}
\caption{Behavior of the (a) diagonal coefficients $|a_n|$ and the off-diagonal coefficients $b_n$ with different dissipative strength for the SYK$_4$ model, with linear dissipators. The dotted line in (a) is given by \eqref{asb}. Our initial operator is $\mathcal{O}_0 = \sqrt{2} \psi_1$, and the number of fermions, $N = 20$. Here we have taken 50 Hamiltonian realizations.} \label{fig:diddSYK1}
\end{figure}

\subsection{For linear jump operators}

\begin{figure}[t]
   \centering
\begin{subfigure}[b]{0.46\textwidth}
\centering
\includegraphics[width=\textwidth]{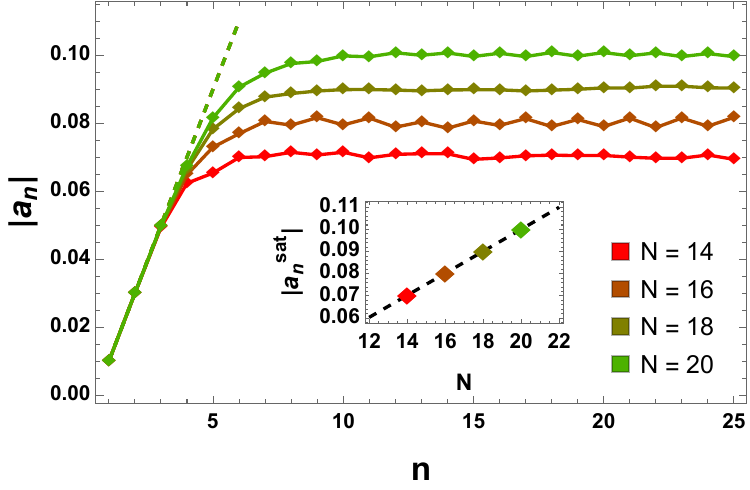}
\caption{Behavior of $|a_n|$.}
\end{subfigure}
\hfill
\begin{subfigure}[b]{0.46\textwidth}
\centering
\includegraphics[width=\textwidth]{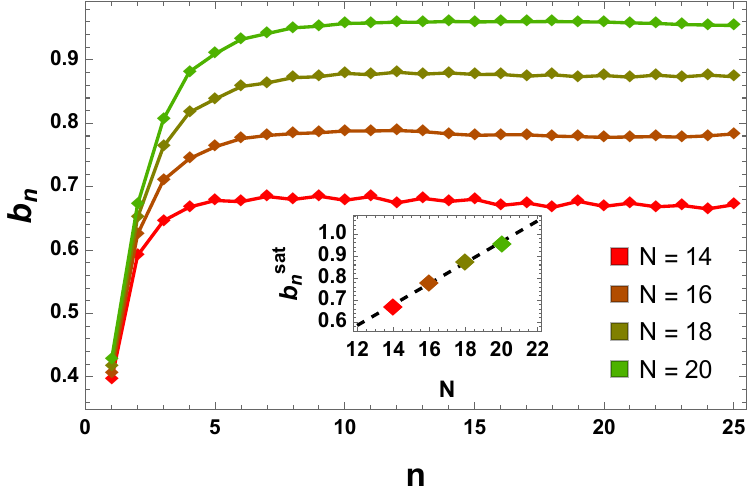}
\caption{Behavior of $b_n$.}
\end{subfigure}
\caption{Behavior of the (a) diagonal coefficients $|a_n|$ and (b) the off-diagonal coefficients $b_n$ with different system sizes for SYK$_4$, with linear dissipators. The insets show the linear dependence (fitted) of the saturation values of $|a_n|$ and $b_n$.  Our initial operator is $\mathcal{O}_0 = \sqrt{2} \psi_1$, and the dissipative strength is fixed at $\lambda = 0.01$. Here we have taken 50 Hamiltonian realizations.} \label{fig:SYK1N2}
\end{figure}

Fig.\,\ref{fig:diddSYK1} shows the result for $N=20$, $\mathrm{SYK}_4$ with $50$ Hamiltonian realizations. We can see that the off-diagonal coefficients are unaffected by the dissipation and they are exactly equal to the closed-system counterparts \cite{Parker:2018yvk}. On the other hand, the dissipation only influences the diagonal coefficients. They are purely imaginary and we can compute the slope of diagonal coefficients as
\begin{align}
    |a_n| = \lambda \,(2n - 1)\,, \label{asb}
\end{align}
 which grows linearly. This slope agrees with the result obtained by Arnoldi iteration in \cite{Bhattacharjee:2022lzy}, except for some constant shift that depends on the intrinsic nature of the algorithm. Given the set of Lanczos coefficients, \textbf{Proposition 1} holds which can be checked explicitly. Moreover, as seen from Fig.\,\ref{fig:SYK1N2}, the slopes of both diagonal and off-diagonal coefficients do not depend on the system size $N$ while their saturation value does. In fact, the saturation linearly increases with the system size $N$, as shown in the insets of Fig.\,\ref{fig:SYK1N2}, i.e.,
 \begin{align}
     |a_n^{\mathrm{sat}}| \propto N\,,~~~ ~b_n^{\mathrm{sat}}
     \propto N\,,
 \end{align}
 for a fixed dissipation strength $\mu$. This finite-size scaling is consistent with previous studies \cite{Jian:2020qpp, Bhattacharjee:2022lzy}. However, in the true thermodynamic limit $N \rightarrow \infty$, we only observe the asymptotic growth \eqref{asymab}.

% However, the Lindbladian matrix becomes an upper Hessenberg matrix in the finite $q$ and finite $N$ limit. 

\subsection{For random quadratic jump operators}

\begin{figure}[t]
   \centering
\begin{subfigure}[b]{0.48\textwidth}
\centering
\includegraphics[width=\textwidth]{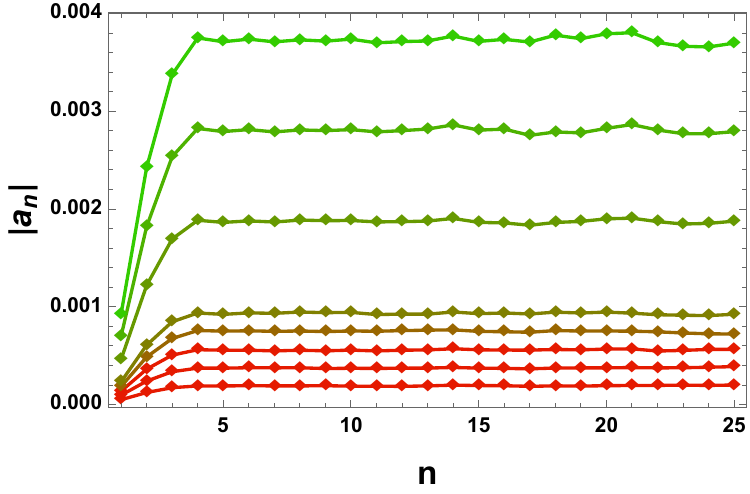}
\caption{Behavior of $|a_n|$.}
\end{subfigure}
\hfill
\begin{subfigure}[b]{0.46\textwidth}
\centering
\includegraphics[width=\textwidth]{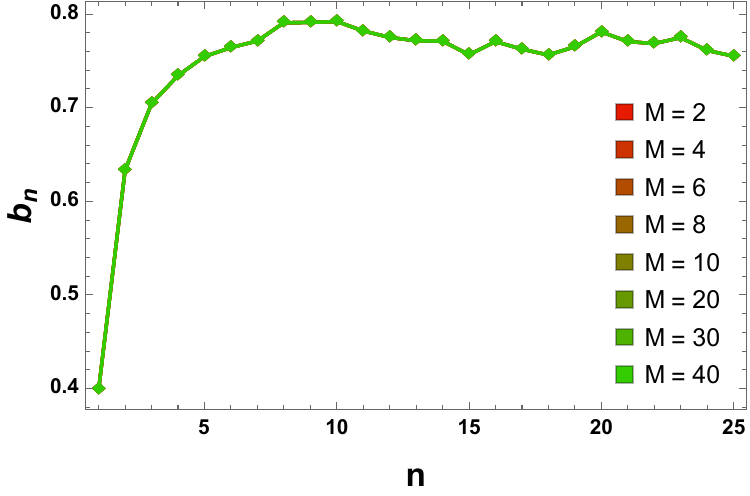}
\caption{Behavior of $b_n$.}
\end{subfigure}
\caption{Behavior of the (a) diagonal coefficients $|a_n|$ and (b) the off-diagonal coefficients $b_n$ with the different number of jump operators $M$ for SYK$_4$. Notice that the $b_n$s exactly overlap for all values of $M$. Our initial operator is $\mathcal{O}_0 = \sqrt{2} \psi_1$, system size is $N =16$ and the dissipative strength is fixed at $V = 0.02$. The random Lindblad operators are taken over $30$ averages.} \label{fig:SYK2N}
\end{figure}

With this choice, we perform the bi-Lanczos algorithm for several numbers of Lindblad operators (i.e., different values of $M$) with a fixed choice of initial operator $\mathcal{O}_0$ and system size $N$. The Lanczos coefficients are shown in Fig.\,\ref{fig:SYK2N}. We see that the diagonal coefficients $|a_n|$ are strongly dependent on the dissipation while the off-diagonal coefficients ($b_n = c_n$) are independent of the dissipation. We also checked that the \textbf{Proposition 1} remains to hold with the observed set of Lanczos coefficients.

\begin{figure}[t]
   \centering
\begin{subfigure}[b]{0.32\textwidth}
\centering
\includegraphics[width=\textwidth]{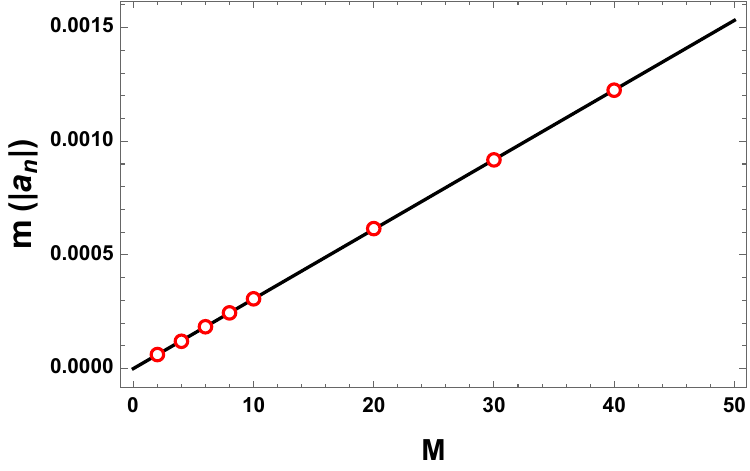}
\caption{Slope of $m (|a_n|)$.} \label{sa}
\end{subfigure}
\hfill
\begin{subfigure}[b]{0.32\textwidth}
\centering
\includegraphics[width=\textwidth]{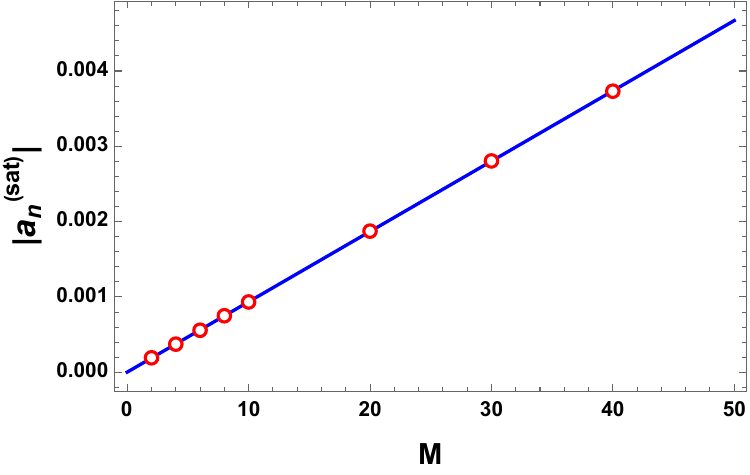} 
\caption{Slope of $|a_n^{\mathrm{(sat)}}|$.} \label{sb}
\end{subfigure}
\hfill
\begin{subfigure}[b]{0.32\textwidth}
\centering
\includegraphics[width=\textwidth]{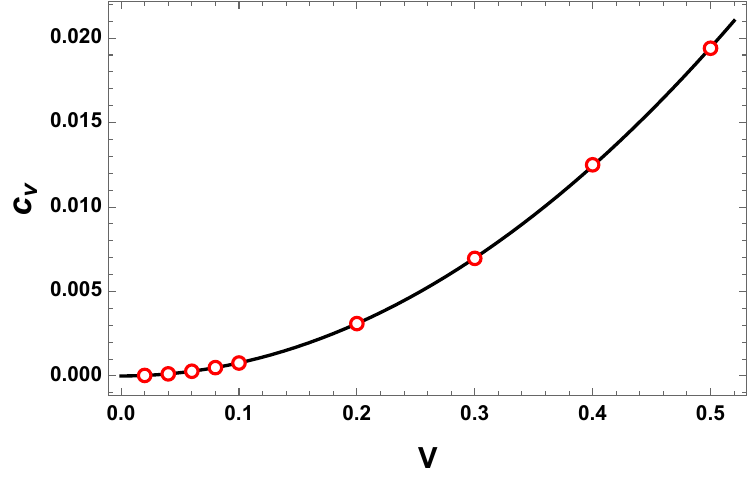}
\caption{$c_V$ with $V$.} \label{sc}
\end{subfigure}
\caption{Behavior of the (a) slope of the diagonal co-efficient $|a_n|$ and (b) the saturation of the diagonal coefficients with the different number of jump operators. Our initial operator is $\mathcal{O}_0 = \sqrt{2} \psi_1$, system size is $N =16$ and the dissipative strength is fixed at $V = 0.02$. The values are obtained after averaging over $30$ disordered realizations. (c) Dependence of $c_V$ with $V$. The numerical fitting gives $\kappa = 0.0780$ and $\beta = 2.0046$ according to \eqref{cv}.} \label{fig:SYK3N}
\end{figure}

We remark on two features of the diagonal coefficients. First, we observe that both the slopes and the saturation values of $|a_n|$ increase with $M$. In Fig.\,\ref{sa} and Fig.\,\ref{sb}, we separately show the behavior of the slope $m (|a_n|)$ and the saturation value $|a_n^{\mathrm{(sat)}}|$ with the number of Lindblad operators. This increase is linear in either case, i.e.,
\begin{align}
    m(|a_n|) \propto M\,, ~~~~\mathrm{and}~~~~ |a_n^{\mathrm{(sat)}}| \propto M\,, ~~~~~ \mathrm{fixed}\, N\,.
\end{align}
From our linear dissipator result, we can also understand that increasing the system size $N$ increases the individual saturation value but does not affect the slope.

Second, we assume that $a_n$ is an asymptotically smooth function of $n$ in the thermodynamic limit. This smoothness behavior is a typical assumption of the operator growth hypothesis for the off-diagonal coefficients \cite{Parker:2018yvk}, although some violations were observed in quantum field theories \cite{Camargo:2022rnt, Avdoshkin:2022xuw}\footnote{In such theories, odd and even coefficients grow linearly with different slopes, mostly controlled by the mass gap of the theory \cite{Camargo:2022rnt} or the compactification radius of the compactified geometry \cite{Avdoshkin:2022xuw}.} However, in this paper, we continue the smoothness assumption which enables us to define the growth rate of the form
\begin{align}
    m(|a_n|) := \frac{d |a_n|}{d n} \propto M \,,
    %~~~~ \Rightarrow ~~~~ \frac{d |a_n|}{d n} \propto R N\,,
\end{align}
with fixed $N$. In the large $N$ and large $q$ limit, the growth will be asymptotic. In other words, we can write
\begin{align}
    a_n \sim i c_{V} M \,n = i c_{V} R N \,n \,. \label{ang}
\end{align}
where the proportionality constant $c_{V}$ depends on the dissipation strength $V$. The second equality comes in a special ``double-scaling limit'', which is defined as $M \rightarrow \infty$ and $N \rightarrow \infty$ keeping $R = M/N$ finite (fixed). Our interest is to find the form of the proportionality constant  $c_{V}$. In principle, it can be either analytically found by computing a two-point function as in \eqref{eq:Ct} or numerically by a fitting of the various data of $c_{V}$. We choose the latter approach.
The numerical data obtained by implementing the bi-Lanczos algorithm suggests the following form
\begin{align}
    c_{V} = \kappa \,V^{\beta}\,, \label{cv}
\end{align}
where $\kappa, \beta$ are some real coefficients and can be obtained by fitting the data which is shown in Fig.\,\ref{sc}. Note that this set is obtained for $N = 16$, and can be improved by increasing $N$. For our interest, the coefficient $\kappa$ is irrelevant and we are primarily interested in the exponent $\beta$. The fitting suggests $\beta = 2.0046 \approx 2$. Hence, we can write \eqref{ang} of the form
\begin{align}
    a_n = i \kappa  R N V^{2} \, n \sim i R V^{2} \,n \,,
\end{align}
This asymptotic growth is consistent with \eqref{ac1} and of the form \eqref{asymab}, while $b_n$ follows the same growth of a closed system with coefficient $\alpha$.

\section{Universal aspects of operator growth in open systems} \label{uni}

In this section, we interpret both the analytic and the numerical results into a concise form and discuss some universal aspects of operator growth, generic to any choice of Lindbladian.

\subsection{An asymptotic growth of Lanczos coefficients and Krylov complexity}

\begin{figure}[t]
   \centering
\includegraphics[width=0.53\textwidth]{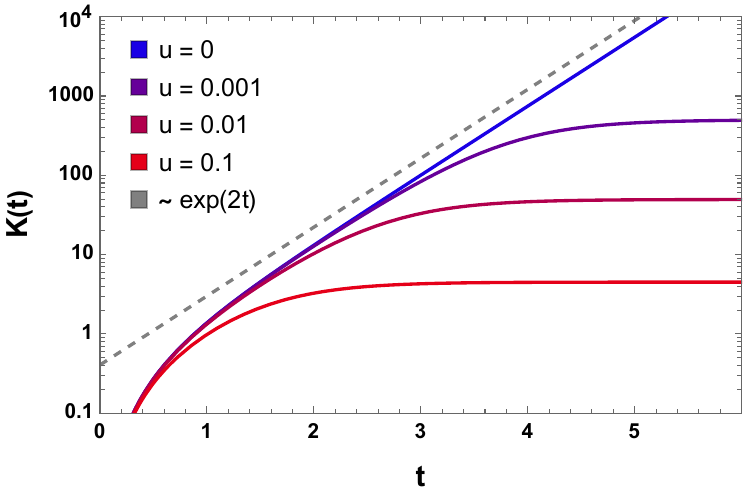}
\caption{The behavior of Krylov complexity \eqref{kcplot} for different dissipation strengths. The gray dashed line indicates the behavior of $\sim e^{2t}$ (we kept some separation from $u=0$ line to have the visual distinguishability) of a closed system. We choose $\eta = 1$.} \label{fig:kcomp}
\end{figure}

The analytical and the numerical analysis motivate us to propose the following sets of asymptotic growth for the Lanczos coefficients in the large-$n$ limit~\cite{Bhattacharjee:2022lzy}
\begin{align}
    a_n \sim i \chi \mu n\,, ~~~~ b_n = c_n \sim \alpha n\,, \label{asymab}
\end{align}
where $\mu$ is the generic dissipative parameter, $\chi$ is some number which is independent of the dissipation and $\alpha$ captures the information of the Hamiltonian and the initial operator. This form motivates an operator growth hypothesis in open quantum systems. As we have from the previous analysis, the growth of $a_n$ is linear, and thus $\mu \propto \lambda$ for the linear dissipator and $\mu \propto R V^2$ for the generic $p$-body dissipator. In either case, the dissipation strength is quadratic due to the simultaneous appearance of $L_k$ and $L_k^{\dagger}$ in the Lindbladian. One can directly calculate the wavefunctions and the K-complexity from this asymptotic growth \eqref{asymab}. First, the Krylov basis wavefunctions $\varphi_{n}$ are given by \cite{Bhattacharjee:2022lzy}
\begin{align}
    \varphi_n(t) =   \frac{ \sech(\gamma t)^\eta }{(1 + u \tanh(\gamma t))^\eta} \,\times 
     (1 - u^2)^{\frac{n}2}  \sqrt{\frac{(\eta)_n}{n!}}    \left( \frac{\tanh (\gamma t)}{1 +u \tanh(\gamma t)} \right)^n \,, \label{kwave}
\end{align}
for the exact expression $b_n^2 = (1-u^2)\gamma^2 n (n-1+\eta)$ and $a_n = i u \gamma (2n + \eta)$ which reduces to \eqref{asymab} for $\alpha^2 = \gamma^2(1-u^2)$ and $\chi \mu = 2 \gamma u$ asymptotically for some $\eta \sim o(1)$. The Krylov complexity can be straightforwardly computed as~\cite{Bhattacharjee:2022lzy}
\begin{align}
    K(t) = \frac{\eta  \left(1-u^2\right) \tanh ^2(\gamma t)}{1+2 u \tanh (\gamma t)-\left(1-2 u^2\right) \tanh ^2(\gamma t)}\,, \label{kcplot}
\end{align}
The behavior of Krylov complexity is shown in Fig.\,\ref{fig:kcomp} for different dissipation strengths. In particular, we observe that the dissipative time scale is logarithmic while the saturation of Krylov complexity scales inversely to the dissipative strength \cite{Bhattacharjee:2022lzy}:
\begin{align}
t_{d} \sim \frac{1}{\gamma}\ln(1/u)\,,~~~~~~ K_{\mathrm{sat}} \sim 1/u\,.
\end{align}
%From our analysis, we can confirm that for any generic dissipator, the dissipative time indeed scales logarithmically with the dissipation
%\begin{align}
%    t_{d} \sim  \begin{cases} \ln(1/\lambda)  \,      & \,p = 1\,,\\
%    \ln[1/(R V^2)]\,, \,   & \,p \geq 2\,,
%    \label{eq:td-main}
%  \end{cases}
%\end{align}
%and the saturation value scales inversely with the dissipation
%\begin{align}
%    K_{\mathrm{sat}} \sim  \begin{cases} 1/\lambda  \,      & \,p = 1\,,\\
%   1/(R V^2)\,, \,   & \,p \geq 2\,,
%    \label{eq:ksat-main1}
%  \end{cases}
%\end{align}
and finally, reach a value that is independent of the system size. This logarithmic timescale and saturation is also found in operator size distribution \cite{Schuster:2022bot}.

%\begin{figure}[t]
%   \centering
%\begin{subfigure}[b]{0.46\textwidth}
%\centering
%\includegraphics[width=\textwidth]{kcomp.pdf}
%\caption{} \label{fig:kcomp}
%\end{subfigure}
%\hfill
%\begin{subfigure}[b]{0.46\textwidth}
%\centering
%\includegraphics[width=\textwidth]{pole.pdf}
%\caption{}
%\end{subfigure}
%\caption{(a) The behavior of Krylov complexity \eqref{kcplot} for different dissipation strengths. The gray dashed line indicates the behavior of $\sim e^{2t}$ (we kept some separation from $u=0$ line to have the visual distinguishability) of a closed system. We choose $\eta = 1$. (b) The pole of the auto-correlation function modifies to the form of \eqref{autopole}.} \label{fig:misc}
%\end{figure}

The scaling of the above saturation invites an interpretation from the quantum measurement. Recall \eqref{st4}, where the rate of measurement is translated as the dissipation strength in the Markovian approximation. In other words, the jump operators can be interpreted as performing a similar task to measurement operators - the environment makes a continuous measurement through it. However, a significant difference from generic measurement is that here the outcome is unknown to us. However, since the measurement is a non-unitary process, the stronger the measurement rate, the lower the probability of the system being evolved by a unitary evolution. In other words, increasing the dissipation strength $\mu$ lowers the possibility of exponential growth which is evident in Fig.\,\ref{fig:kcomp}.

%\begin{figure}[t]
%   \centering
%\begin{subfigure}[b]{0.46\textwidth}
%\centering
%\includegraphics[width=\textwidth]{kcomp.pdf}
%\caption{} \label{fig:kcomp}
%\end{subfigure}
%\hfill
%\begin{subfigure}[b]{0.46\textwidth}
%\centering
%\includegraphics[width=\textwidth]{sigcomp.pdf}
%\caption{} \label{fig:sigcomp}
%\end{subfigure}
%\caption{The behavior of (a) Krylov complexity \eqref{kcplot} and (b) signal-to-noise-ratio for different dissipation strengths. We choose $\eta = 1$.} \label{fig:kcompp}
%\end{figure}

%We can also compute the higher moments, namely the Krylov variance defined as~\cite{Bhattacharjee:2022lzy}
%\begin{align}
%    \delta_K (t) := \frac{1}{\mathcal{Z}} \sum_n (n-K(t))^2 =  \frac{\eta  \left(1-u^2\right) \tanh ^2(t) (1+ u \tanh (t))^2}{(1+2 u \tanh (t)-\left(1-2 u^2\right) \tanh ^2(t))^2}\,, \label{dcplot}
%\end{align}
%At early time, it behaves exponentially $\delta_K (t) \sim e^{4 t}$ while the late-time saturation scales as $\delta_K^{\mathrm{sat}} \sim 1/4 u^2$. The signal-to-noise ratio is defined as $\delta K(t) := \sqrt{\delta_K(t)}/K(t)$. The behavior is shown in Fig.\,\ref{fig:sigcomp}, approaching $\delta K(t) \sim 1 + u + o(u^2)$ at late times. 

%Hence, we can directly advocate the asymptotic analysis of \cite{Bhattacharjee:2022lzy} which gives
%\begin{align}
%    t_{d} \sim \ln\left(\frac{1}{V}\right)\,,~~~~~~ K_{\mathrm{sat}} \sim \frac{1}{V^2}\,.
%\end{align}
%The ``dissipative time scale'' is logarithmic to the dissipative strength while the saturation value of K-complexity is inversely proportional to it.

\begin{figure}[t]
   \centering
\includegraphics[width=0.68\textwidth]{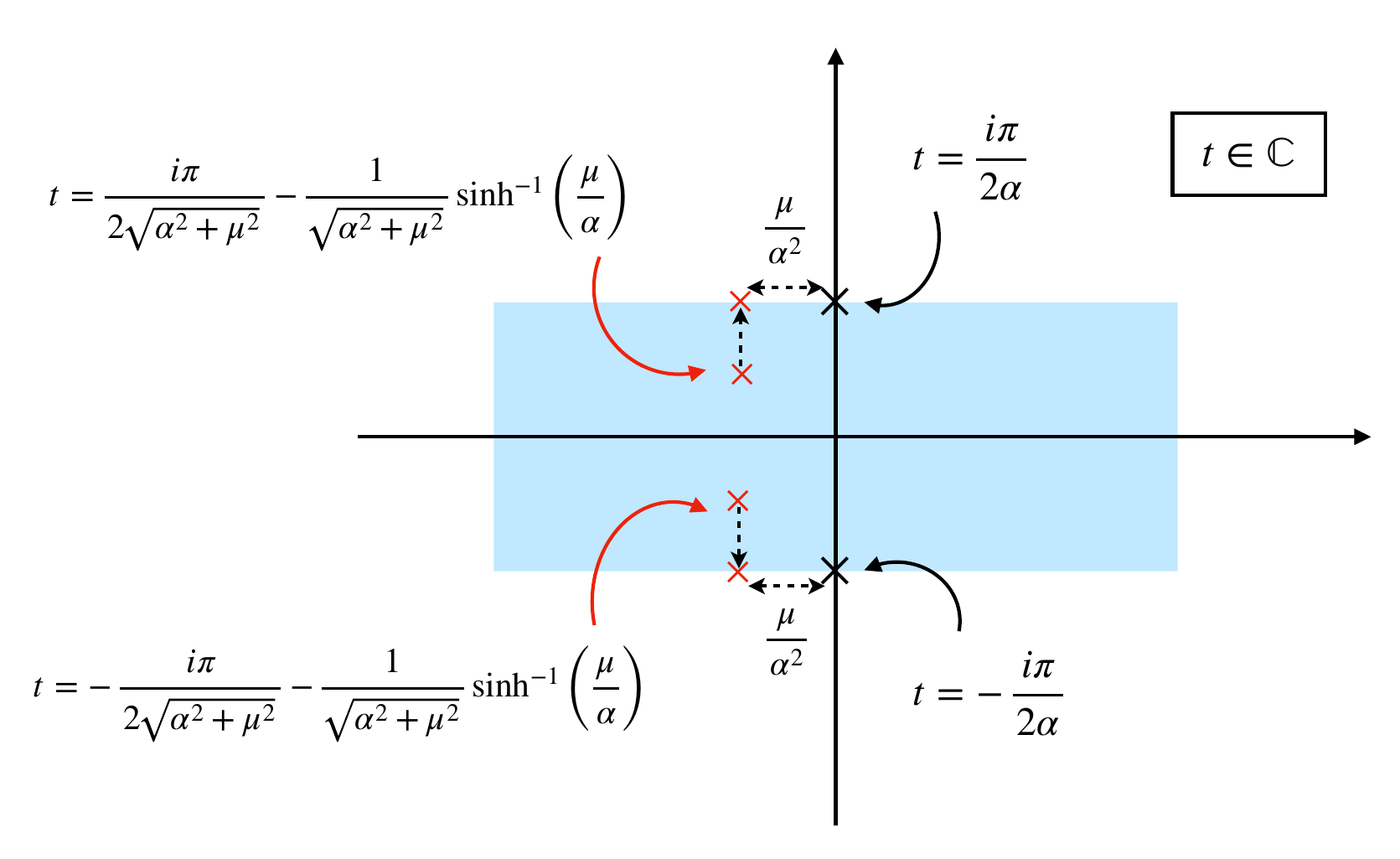}
\caption{The pole of the auto-correlation function in the complex $t$ plane. The black crosses indicate the poles without dissipation i.e., $\mu =0$ or the leading order term in \eqref{mmo}, with the blue-shaded region as the region of analyticity. The red crosses indicate the poles with auto-correlation of the form of \eqref{autopole}. In the weak coupling regime, the poles move away from each other along the $y$ axis, thus resulting in an effect only a linear shift with magnitude $\mu/\alpha^2$.} \label{fig:pole}
\end{figure}

\subsection{Pole structures of auto-correlation and spectral function}

The two sets of Lanczos coefficients generically modify the behavior of the auto-correlation function and correspondingly its Fourier transform, known as the spectral function. It is interesting to investigate the pole structure of the auto-correlation function, similar to its closed system counterpart \cite{Parker:2018yvk}. In particular, our above analysis suggests that we might devise a generic form of an auto-correlation function. Let us assume a form of the auto-correlation function
\begin{align}
    C(\mu,t) = \frac{1}{\alpha}\sqrt{\alpha^2+\mu^2}\,\text{sech}\left( t \sqrt{\alpha^2 + \mu^2} + \sinh^{-1}(\mu/\alpha) \right)\,, \label{poleauto}
\end{align}
with $\alpha$ being some constant which is independent of $\mu$. This form of the autocorrelation function looks similar to \eqref{eq:Ct}-\eqref{ect1}.\footnote{One can, in principle consider a variety of possible another set of functions that satisfy the required properties of an auto-correlation function. For example, we can add a functional form $f(\mu, t)$ with the property $f(0,t) = f(\mu,0) = 0$. This will alter the asymptotic growth of the Lanczos coefficients. However, since we are considering a particular prescribed set of coefficients, we take the form of \eqref{poleauto}.} We can easily see $C(0,t)= \text{sech}(\alpha t)$ reduces to the known closed system counterpart \cite{Parker:2018yvk, Barbon:2019wsy} and $C(\mu,0) = 1$. In other words, we can take this as a two-variable function and forget about its origin. Thus it is valid for any $\mu$, not necessarily small. In fact, this function gives an asymptotic linear behavior of both $a_n = i \alpha \mu (2n+1) \sim i \chi \mu n$  and $b_n = \alpha n$ of the form \eqref{asymab}, without making any approximation of $\mu$. Note that for $\mu=0$, the auto-correlation \eqref{poleauto} reduces to $C(0,t) = \sech(\alpha t)$, which is obtained for $a_n =0$ and $b_n = \alpha n$ \cite{Parker:2018yvk, Barbon:2019wsy}. To investigate the pole structure of \eqref{poleauto}, we set it to zero and find that the closest pole is located as
\begin{align}
    t_{\pm} = \pm \frac{i \pi}{2\sqrt{ \alpha^2 +\mu^2}} - \frac{1}{ \sqrt{ \alpha^2 +\mu^2}}  \sinh^{-1}\left(\frac{\mu}{\alpha}\right)\,. \label{autopole}
\end{align}
The pole is not exactly at the imaginary $t$ axis, rather is it shifted (see Fig.\,\ref{fig:pole}). Then a reasonable question is to ask which kind of system has such a form of auto-correlation? Our answer is the dissipative SYK in the weak coupling regime, modeled by any generic random $p$-body Lindblad operators. Then our $\alpha$ dictates the system strength while $\mu$ encodes the dissipative strength. We have already found such results both analytically and numerically in previous sections. Only then, does the pole structure of \eqref{autopole} give the information of the operator growth. Thus, to connect with such growth, we expand \eqref{autopole} in the small $\mu$ regime as
\begin{align}
     t_{\pm} = \pm \frac{i \pi}{2 \alpha} - \frac{\mu}{ \alpha^2} + o(\mu^2)\,. \label{mmo}
\end{align}
The $o(\mu^2)$ terms do contain both real and imaginary parts but they are not relevant to our discussion. In fact, \eqref{autopole} does suggest that the pole is not only squeezed by a factor of $\sqrt{\alpha^2+ \mu^2}$ along the imaginary $t$ axis but also shifted by a length of $\sinh^{-1}(\mu/\alpha)/( \sqrt{\alpha^2+\mu^2}$ on the direction of negative $x$ axis. The combined effect has a diagonal shift (see Fig.\,\ref{fig:pole}), squeezing the poles into the domain of the analyticity of \eqref{poleauto} for $\mu =0$.  However, in the weak dissipation regime, the squeezing is no longer required in the leading order and the closest pole structure still gives the growth of $b_n$ and thus the Krylov exponent. The effect of dissipation merely affects a linear shift of the pole. Of course, at zero dissipation, the poles are exactly located at $t = \pm i\pi/(2\alpha)$ \cite{Parker:2018yvk}.

\begin{figure}[t]
   \centering
\includegraphics[width=0.6\textwidth]{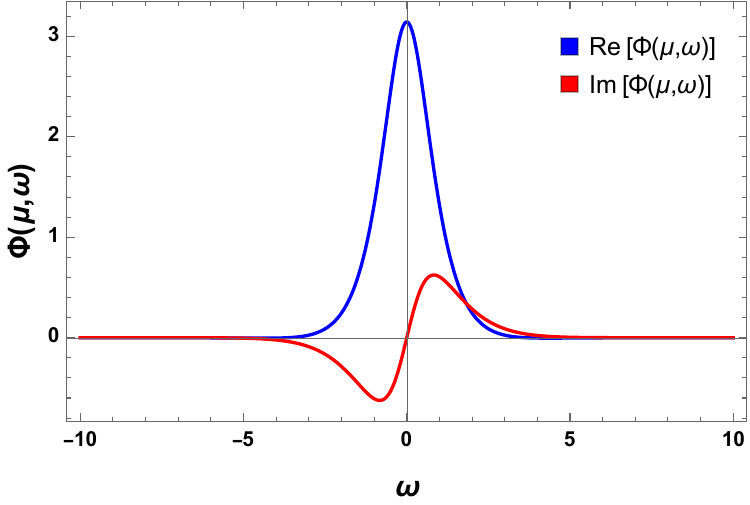}
\caption{The real and the imaginary part of the spectral function \eqref{mufreqBB} for $\mu =0.5$. We choose $\alpha =1$.} \label{fig:sp}
\end{figure}

The spectral function is given by the Fourier transform of the auto-correlation function, i.e.,
\begin{align}
    \Phi(\mu,\omega) = \int_{-\infty}^{\infty} e^{- i \omega t} \,C(\mu,t)\,,
\end{align}
Taking the auto-correlation of the form \eqref{poleauto}, we find
\begin{align}
  \Phi(\mu, \omega) &=    \frac{\pi}{\alpha} \text{sech}\left(\frac{\pi  \omega }{2 \sqrt{\alpha^2+\mu^2}}\right) e^{\frac{i \omega }{\sqrt{\alpha^2+\mu^2}} \sinh^{-1}\left(\frac{\mu }{\alpha }\right)} \,.\label{mufreqBB}
\end{align}
This is also a completely generic result, for any $\alpha$, $\mu$, and $\omega$. However, we want to connect it to the operator growth and thus in the weak dissipation regime, Eq.\,\eqref{mufreqBB} becomes
\begin{align}
    \Phi(\mu, \omega)\Big|_{\mu \rightarrow 0} &= \frac{\pi}{\alpha} \sech\left(\frac{\pi \omega}{2 \alpha}\right) + \frac{i\pi \omega \mu}{\alpha^3} \sech\left(\frac{\pi \omega}{2 \alpha}\right) + o(\mu^2)\, \label{ldn} \\
    &=  \left(1 + i \frac{\omega \mu}{\alpha^2}\right) \frac{\pi}{\alpha} \sech\left(\frac{\pi \omega}{2 \alpha}\right) + o(\mu^2) \,.  \label{ldn2}
\end{align}
The leading term in \eqref{ldn} is the known result for $\mu =0$ as
\begin{align}
    \Phi(0,\omega) = \frac{\pi}{\alpha} \sech\left(\frac{\pi \omega}{2 \alpha}\right) ~ \sim ~\frac{2 \pi}{\alpha} e^{-\frac{\pi |\omega|}{2 \alpha}}\, ~~~~~ \mathrm{for~large~}\omega\,, \label{zerofreq}
\end{align}
which shows a long exponential tail in the large frequency regime \cite{Parker:2018yvk}. The subleading term in \eqref{ldn2} depends on the ratio $\mu/\alpha$, and as long as $\mu \ll \alpha$, the leading term dominates. Note that the leading term decays as $\exp(-\#|\omega|)$, while the subleading term decays as $\omega \exp(-\#|\omega|)$. The overall decay still follows the leading behavior for large $\omega$. Our analysis assumes a smooth behavior of the Lanczos coefficients unlike \cite{Avdoshkin:2019trj, Avdoshkin:2022xuw}, where the different behavior of even and odd coefficients may lead to incorrect consequences on the spectral function.\footnote{We thank Anatoly Dymarksy for pointing out this to us.} Thus, in the weak dissipation regime, the linear shift of the pole in the auto-correlation function is sublinear in $\mu$, which reflects an $\omega \exp(-\#|\omega|)$ decay in the sublinear term of the spectral function. This posits an alternate form of the operator growth hypothesis in open quantum systems.

\section{Lindbladian dynamics: OTOC and $\mathfrak{q}$-complexity} \label{otocsec}
In this section, we derive an analytic expression for the out-of-time-order correlator (OTOC) for a time-evolved $p$-body fermionic operator and some general aspects of $\mathfrak{q}$-complexity.
\subsection{Lindbladian dynamics for OTOC}
We start with the OTOC of a single-body fermionic operator. The expression to be evaluated is the following
\begin{align}
    \mathrm{OTOC}(t) = \frac{1}{2\mathcal{Z}(t) N}\sum_{j = 1}^{N}\text{Tr}[\rho_\infty \{\psi_{j},\psi_{1}(t)\}\{\psi_{j},\psi_{1}(t)\}^\dagger]\,,
\end{align}
where $\mathcal{Z}(t) =\text{Tr}[\rho_\infty \psi_{1}(t)^\dagger\psi_{1}(t)]$ is the normalisation factor and $\rho_\infty = \frac{1}{2^{N/2}}\mathbf{I}$. The overall factor of $1/2$ arises because we use unnormalized $\psi_{j}$, and the $1/N$ factor is introduced to account for averaging over the full set of Majorana fermions \cite{Roberts:2018mnp}. The key idea is simple: the knowledge of the Krylov basis of $\psi_1$ allows us to write the time-evolved operator $\psi_{1}(t)$ in Krylov basis as follows
\begin{align}
    \psi_{1}(t) = \frac{1}{\sqrt{2}} \sum_{k}i^{k} \varphi_{k}(t)\mathcal{O}_{k}\,\label{inopotoc}.
\end{align}
In general, this would not give us a lot of analytical prowess. In the limit of large $q$ and large $N$, however, things become tractable due to operator concentration \eqref{Onterm}. Recall that the Krylov vectors are naturally orthonormal to each other. Using this we can evaluate the expression for the commutator $\{\psi_{j},\psi_{1}(t)\}$ as 
\begin{align}
    \{\psi_{j},\psi_{1}(t)\} = \frac{1}{\sqrt{2}} \sum_{k}i^{k} \varphi_{k}(t)\{\psi_{j}, \mathcal{O}_{k}\}\,,
\end{align}
A general property of Krylov vectors is that basis vectors generated from a Hermitian Hamiltonian and initial operator are alternatingly Hermitian and anti-Hermitian. This allows us to evaluate the Hermitian conjugate of the above expression quite simply by noting that the combination $i^{k}\mathcal{O}_{k}$ is Hermitian. In other words,
\begin{align}
    \{\psi_{j},\psi_{1}(t)\}^\dagger = \frac{1}{\sqrt{2}} \sum_{k}i^{k}\varphi^{*}_{k}(t)\{\psi_{j}, \mathcal{O}_{k}\}\,.
\end{align}
The trace operation under the sum is written as
\begin{align}
    \text{Tr}[\rho_\infty \{\psi_{j},\psi_{1}(t)\}\{\psi_{j},\psi_{1}(t)\}^\dagger] = \frac{1}{2^{N/2 + 1}}\sum_{k = 1}\sum_{l = 1}i^{k + l}\varphi_{k}(t)\varphi^{*}_{l}(t)\text{Tr}[\{\psi_{j},\mathcal{O}_{k}\}\{\psi_{j},\mathcal{O}_{l}\}]\,,
\end{align}
where we have used the fact that $\rho_\infty = \frac{1}{2^{N/2}}\mathbf{I}$. The final task is to evaluate the trace operation in the RHS of the expression above, i.e.,
\begin{align}
    \text{Tr}[\{\psi_{j},\mathcal{O}_{k}\}\{\psi_{j},\mathcal{O}_{l}\}] = \text{Tr}[\mathcal{O}_{k}\mathcal{O}_{l}] + 2 \text{Tr}[\psi_j \mathcal{O}_{k} \psi_{j} \mathcal{O}_{l}]\,.
\end{align}
Note that the first term only contributes if $k = l$ and (assuming that the Krylov vectors are properly normalized) contributes $(-1)^{k}2^{N/2}$. The second term is the one that has to be evaluated carefully. We note that the trace of the product of fermion strings of different lengths vanishes. This implies that we have a non-zero contribution coming from $k = l$ only. Therefore the full expression becomes
\begin{align}
    \text{Tr}[\{\psi_{j},\mathcal{O}_{k}\}\{\psi_{j},\mathcal{O}_{l}\}] = (-1)^{k}2^{N/2}\delta_{k l} + 2 \text{Tr}[\psi_j \mathcal{O}_{k} \psi_{j} \mathcal{O}_{k}]\delta_{k l}\,.
\end{align}
We now look at the term $\psi_{j}\mathcal{O}_{k}$ in the second trace. It is straightforward to see that the following result is true
\begin{align}
    \psi_{j}\sum_{1 \leq i_1 < i_2 < \dots < i_{s} \leq N}c_{i_{1} i_{2} \dots i_{s}}\psis = \sum_{1 \leq i_1 < i_2 < \dots < i_{s} \leq N}c_{i_{1} i_{2} \dots i_{s}} (-1)^{\sum_{l = 1}^{s}\delta_{i_{l}, j} + s}\psis\psi_{j}\,.
\end{align}
Therefore the second trace term becomes
\begin{align}
    \text{Tr}[\psi_j \mathcal{O}_{k} \psi_{j} \mathcal{O}_{k}] = \frac{(-1)^{k + s}}{2^{s + 1 - \frac{N}{2}}}\sum_{1 \leq i_1 < i_2 < \dots < i_{s} \leq N}\vert c_{i_{1} i_{2} \dots i_{s}}\vert^2 (-1)^{\sum_{l = 1}^{s}\delta_{i_{l}, j}}\,,
\end{align}
where in the last step we have used the fact that $c^{*}_{i_{1} i_{2} \dots i_{s}} = (-1)^{k q/2}c_{i_{1} i_{2} \dots i_{s}}$ in order to ensure that $i^{k}\mathcal{O}_{k}$ is Hermitian. Combining the two terms we get the following expression
\begin{align}
    \text{Tr}[\{\psi_{j},\mathcal{O}_{k}\}\{\psi_{j},\mathcal{O}_{l}\}] = (-1)^{k}2^{N/2}\left(1 - \frac{(-1)^{s}}{2^{s}}\sum_{1 \leq i_1 < i_2 < \dots < i_{s} \leq N}\vert c_{i_{1} i_{2} \dots i_{s}}\vert^2 (-1)^{\sum_{l = 1}^{s}\delta_{i_{l}, j}}\right)\,.
\end{align}
Utilizing this, the trace operation under the full sum becomes
\begin{align*}
     \text{Tr}[\rho_\infty \{\psi_{j},\psi_{1}(t)\}\{\psi_{j},\psi_{1}(t)\}^\dagger] = \frac{1}{2}\sum_{k}\vert\varphi_{k}(t)\vert^2\left(1 + \frac{(-1)^{s}}{2^{s}}\sum_{1 \leq i_1 < i_2 < \dots < i_{s} \leq N}\vert c_{i_{1} i_{2} \dots i_{s}}\vert^2 (-1)^{\sum_{l = 1}^{s}\delta_{i_{l}, j}}\right)\,.
\end{align*}
We also note that $\mathcal{Z} = \text{Tr}[\rho_\infty \psi_{1}^\dagger(t) \psi_{1}(t)] = \frac{1}{2}\sum_{k}\vert\varphi_{k}(t)\vert^2$. The last piece of the ingredient is noting that since the Krylov basis vectors are normalized, it follows that $\sum_{1 < i_1 < i_2 < \dots < i_{s} \leq N}\vert c_{i_{1} i_{2} \dots i_{s}}\vert^2 = 2^{s}$. From these pieces of information, we can write the following expression for the OTOC
\begin{align}
    \mathrm{OTOC}(t) = \frac{1}{2 N\sum_{k}\vert\varphi_{k}(t)\vert^2}\sum_{j = 1}^{N}\sum_{k}\vert\varphi_{k}(t)\vert^2\left(1 + \frac{(-1)^{s}}{2^{s}}\sum_{1 \leq i_1 < i_2 < \dots < i_{s} \leq N}\vert c_{i_{1} i_{2} \dots i_{s}}\vert^2 (-1)^{\sum_{l = 1}^{s}\delta_{i_{l}, j}}\right)\,. \label{otoceqn1}
\end{align}
There is a natural bound on this quantity, which follows from the fact that the sum of the coefficients in the expression satisfy $\frac{1}{2^{s}}\sum_{1 \leq i_1 < i_2 < \dots < i_{s} \leq N}\vert c_{i_{1} i_{2} \dots i_{s}}\vert^2 (-1)^{\sum_{l = 1}^{s}\delta_{i_{l}, j}} \leq 1$.
\begin{align}
    0 \leq \mathrm{OTOC}(t) \leq 1\,.
\end{align}
In order evaluate the expression \eqref{otoceqn1} further, we evaluate the following sum 
\begin{align}
    \sum_{1 \leq i_1 < i_2 < \dots < i_{s} \leq N}(-1)^{\sum_{l = 1}^{s}\delta_{i_{l}, j}} =  \sum_{1 \leq i_1 < i_2 < \dots < i_{s} \leq N,\, j \notin \{i_{s}\}}1 +  \sum_{1 \leq i_1 < i_2 < \dots < i_{s} \leq N,\, j \in \{i_{s}\}}1\,.
\end{align}
The first sum is simple. It involves performing the sum over the indices $\{i_{1},\dots,i_{s}\}$ over $N - 1$ values (i.e. excluding $j$). This evaluates to
\begin{align}
    \sum_{1 \leq i_1 < i_2 < \dots < i_{s} \leq N,\, j \notin \{i_{s}\}}1 = \binom{N - 1}{s}\,.
\end{align}
The second sum is non-trivial. To evaluate this we consider the cases where $j \in \{i_{1},\dots,i_{s}\}$. This is broken up into two pieces, corresponding to $j < s$ and $j \geq s$. For $j \geq s$, any $i_{m}$ can be chosen from $i_{1},\dots,i_{s}$ and set equal to $j$. Once the $i_{m}$ is chosen, the sum breaks into two pieces.
\begin{align}
    \sum_{1 \leq i_1 < i_2 < \dots < i_{s} \leq N,\, j  \in \{i_{s}\} \geq s}1 &= \sum_{m = 1}^{s}\left(\sum_{1 \leq i_{1} < i_2 < \dots , i_{m - 1} < j}1\right)\left(\sum_{j < i_{m + 1} < i_2 < \dots , i_{s} \leq N}1\right)\notag\\
    &= \sum_{m = 1}^{s}\binom{j - 1}{m - 1}\binom{N - j}{s - m} = \binom{N - 1}{s - 1}\,.
\end{align}
The remaining sum is 
\begin{align}
    \sum_{1 \leq i_1 < i_2 < \dots < i_{s} \leq N,\, j  \in \{i_{s}\} < s}1 = \sum_{m = 1}^{j}\left(\sum_{1 \leq i_{1} < i_2 < \dots , i_{m - 1} < j}1\right)\left(\sum_{j < i_{m + 1} < i_2 < \dots , i_{s} \leq N}1\right)
    = \binom{N - 1}{s - 1}\,.
\end{align}
Therefore in both these cases, the sums evaluate to the same value. Thus the full sum is 
\begin{align}
     \sum_{1 \leq i_1 < i_2 < \dots < i_{s} \leq N}(-1)^{\sum_{l = 1}^{s}\delta_{i_{l}, j}} = \binom{N - 1}{s} - \binom{N - 1}{s - 1}\,. \label{otocoddsum}
\end{align}
Now, we note that under disorder averaging, all the coefficients $\vert c_{i_{1}i_{2}\dots i_{s}}\vert^2$ must be equal. Given that their sum is $2^{s}$, each individual term is equal to 
\begin{align}
    \vert c_{i_{1}i_{2}\dots i_{s}}\vert^2 = 2^{s} \,\binom{N}{s}^{-1}\,. \label{otoccoeffnorm}
\end{align}
Inserting \eqref{otocoddsum} and \eqref{otoccoeffnorm} in \eqref{otoceqn1}, we obtain the following expression
\begin{align}
    \mathrm{OTOC}(t) &=  \frac{1}{2 N\sum_{k}\vert\varphi_{k}(t)\vert^2}\sum_{j = 1}^{N}\sum_{k}\vert\varphi_{k}(t)\vert^2\left(1 + \frac{(-1)^{s}}{2^{s}}\frac{2^{s}}{\binom{N}{s}}\left( \binom{N - 1}{s} - \binom{N - 1}{s - 1}\right)\right) \notag\\
    &= \frac{1}{2 N\sum_{k}\vert\varphi_{k}(t)\vert^2}\sum_{j = 1}^{N}\sum_{k}\vert\varphi_{k}(t)\vert^2\left(1 + \frac{(-1)^{s}}{\binom{N}{s}}\left( \binom{N - 1}{s} - \binom{N - 1}{s - 1}\right)\right) \notag\\
    &= \frac{1}{\sum_{k}\vert\varphi_{k}(t)\vert^2}\sum_{k}\vert\varphi_{k}(t)\vert^2\left( \frac{k(q-2) + 1}{N}\right)\,,\label{sotoc}
\end{align}
where in the final step we have inserted the expression for $s$ in terms of $k$, noting that it is odd for even $q$. Therefore an analytic expression for the OTOC is given by
\begin{align}
    \mathrm{OTOC}(t) =  \frac{1}{\sum_{k}\vert\varphi_{k}(t)\vert^2}\sum_{k}\vert\varphi_{k}(t)\vert^2\left(\frac{k(q-2) + 1}{N}\right) = \frac{\langle s \rangle}{N}\,. \label{otoceqn2}
\end{align}
We note that for the Lanczos coefficients endowed with the fairly generic form $b_{n}^2 = (1 - u^2)n(n - 1 + \eta)$, $a_{n} = i u (2 n + \eta)$, the Krylov basis wavefunctions $\varphi_{n}$ are given by \eqref{kwave}. Using this in \eqref{otoceqn2}, we obtain the following expression
\begin{align}
    &\mathrm{OTOC}(t) = \frac{\tanh ^2(t) \left(\eta  (q-2) \left(1-u^2\right)+2 u^2-1\right)+2 u \tanh (t)+1}{N+N \tanh (t) \left(\left(2 u^2-1\right) \tanh (t)+2 u\right)}\,. \label{ootc}
\end{align}
\begin{figure}[t]
   \centering
\includegraphics[width=0.53\textwidth]{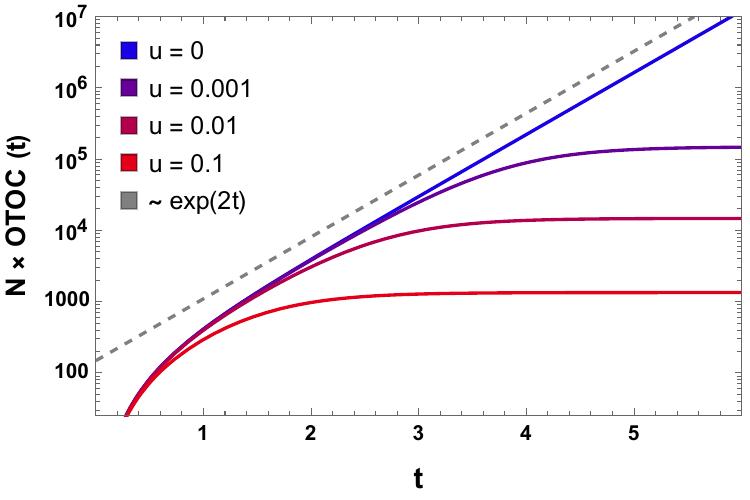}
\caption{The behavior of OTOC \eqref{ootc} for different dissipation strengths. The gray dashed line indicates the behavior of $\sim e^{2t}$ (we kept some separation from $u=0$ line to have the visual distinguishability) of a closed system. We choose $\eta = 1$ and $q=300$. The behavior is similar to \eqref{fig:kcomp}.} \label{fig:otoc}
\end{figure}
Fig.\,\ref{fig:otoc} shows the behavior of OTOC with different dissipative strengths. At late-times ($t \rightarrow \infty$), the OTOC saturates to 
\begin{align}
    \mathrm{OTOC}(t)\big|_{t \rightarrow \infty} = \frac{1}{N}\left(1+\frac{\eta  (q-2) (1-u)}{2 u}\right)\,.
\end{align}
A useful estimate of the saturation timescale is the time at which the saturation value of the OTOC for a given $u$ is equal to the $u \rightarrow 0$ limit of the OTOC. It is straightforward to see that this timescale is given by
%\begin{align}
%    t_{*} \sim \tanh^{-1}\left(\sqrt{\frac{q(1 - u)}{q + (q-4)u}}\right) \xrightarrow[]{u \rightarrow 0} \frac{1}{2}\ln\left(\frac{2 q}{(q-2)u}\right)
%\end{align}
\begin{align}
    t_{*} = \tanh^{-1}\left(\sqrt{\frac{q(1 - u)}{q + (q-4)u}}\right) ~\sim~ \frac{1}{2}\ln\left(\frac{2 q}{(q-2)u}\right)\,, ~~~~ u \rightarrow 0\,.
\end{align}
This timescale is logarithmic in terms of the inverse dissipation strength.

%\BB{The general initial body OTOC: Incomplete for now}

The analysis so far has been considered for the time-evolved initial operator $\psi_{1}$. It is straightforward to see that exactly the same result holds for any single-body initial operator $\psi_{i}$. However, the same results will not hold exactly for the general $p$-body initial operator of the form $\psi_{i_1}\psi_{i_2}\dots \psi_{i_p}$ with $i_{1} < i_{2} < \dots < i_{p}$. As we have discussed, operator concentration holds for any initial operator string. Therefore, most of the discussion will follow through with minor modifications (like replacing $1/\sqrt{2}$ in \eqref{inopotoc} by $1/2^{p/2}$). One subtlety to be pointed out is that this $c^{*}_{i_{1} i_{2} \dots i_{s}} = (-1)^{(k q - p(p-1))/2}c_{i_{1} i_{2} \dots i_{s}}$ for a $p$-body initial operator. The normalisation factor is now $\mathcal{Z} = \text{Tr}[\rho_\infty \mathcal{O}^\dagger\mathcal{O}] = 1/2^p$. This cancels the factor of $1/2^{p/2}$ mentioned before.

The remainder of the calculation described above goes through in an identical fashion. The difference occurs at the final step of \eqref{sotoc}. The OTOC is there given by
\begin{align}
    \mathrm{OTOC}(t) &= \frac{1}{2 N\sum_{k}\vert\varphi_{k}(t)\vert^2}\sum_{j = 1}^{N}\sum_{k}\vert\varphi_{k}(t)\vert^2\left(1 + \frac{(-1)^{s}}{\binom{N}{s}}\left( \binom{N - 1}{s} - \binom{N - 1}{s - 1}\right)\right) \notag\\
    &= \frac{1}{2\sum_{k}\vert\varphi_{k}(t)\vert^2}\sum_{k}\vert\varphi_{k}(t)\vert^2\left( 1 + (-1)^{s}(1 - \frac{2 s}{N})\right)\,,\label{sotoc2}
\end{align}
When $s$ is odd (i.e. if $p$ is odd), the term in the bracket becomes $2s/N$, and for even $s$ (i.e. for even $p$) the term becomes $(2  - 2 s)/N$. However, note that the term $1/N$ just contributes a constant $1/N$. Thus we are left with 
\begin{align}
    \mathrm{OTOC}(t) = \frac{(-1)^{p + 1}\langle s \rangle}{N} + \frac{1 + (-1)^{p}}{2 N}\,.
\end{align}
This concludes the discussion on OTOCs. 
%\textcolor{red}{PN: Can you write few lines of $\mathfrak{q}$-complexity at the end of this section?  This is just because we have mentioned the term $\mathfrak{q}$-complexity in the abstract.}
%\BB{BB: Yes, will do. I have derived a general expression for the $\mathfrak{q}-$ complexity, though it is not useful. Will write that.}

%\BB{$\mathfrak{q}$-complexity : Incomplete for now} 
\subsection{Aspects of $\mathfrak{q}$-complexity}
We briefly discuss the formal results expected for $\mathfrak{q}$-complexity. It is a generalization of the operator growth probes like Krylov complexity, OTOC, and operator size. The notion is defined in detail in \cite{Parker:2018yvk}. We briefly describe the same below :
\begin{itemize}
    \item For a positive semi-definite superoperator $\mathcal{Q}$, one can write it in its' own eigenbasis as follows
    \begin{align}
        \mathcal{Q} = \sum_{i}q_{i}\vert q_{i} )( q_{i} \vert\,.
    \end{align}
    \item There exists a number $K$ such that
    \begin{align}
        (q_{i}\vert \mathcal{L} \vert q_{j}) = 0 \;\;\;\; \forall \; ~~\vert q_i - q_j \vert > K \,.
    \end{align}
    This condition ensures that the expectation value of $\mathcal{Q}$ does not change massively under one application of the Liouvillian.
    \item Similarly, there exists some number $K'$ for an initial operator $\mathcal{O}$ such that
    \begin{align}
        (q_{i}\vert \mathcal{O}) = 0 \;\;\;\; \forall \; ~~\vert q_{i} \vert > K' \,.
    \end{align}
    This condition ensures that the initial operator has a low value of the complexity.
\end{itemize}
The $\mathfrak{q}$-complexity of an initial operator $\mathcal{O}$ is then defined as 
\begin{align}
    \mathcal{Q}(t) = (\mathcal{O}(t) \vert \mathcal{Q} \vert \mathcal{O}(t) )\,.
\end{align}
For our purposes, in the large $-N$ large $-q$ SYK model, it suffices to realize that any general superoperator in the space of Majorana fermions can be written as
\begin{align}
    \mathcal{Q} = \sum_{s, t}\sum_{i_{1},\dots,i_{s},j_{1},\dots,j_{t}}q^{i_{1},\dots,i_{s}}_{j_{1},\dots,j_{t}}\,\vert \psi_{i_1}\dots\psi_{i_s})(\psi_{j_1}\dots\psi_{j_t}\vert\,.
\end{align}
Due to operator concentration, one can write the time-evolved operator $\mathcal{O}(t)$ as follows
\begin{align}
    \vert\mathcal{O}(t)) = \sum_{k}i^{k}\varphi_{k}(t)\vert\mathcal{O}_{k})\,,
\end{align}
where $\vert\mathcal{O}_{k}) = \sum_{1 \leq i_{1} < \dots < i_{d} \leq N}c_{i_{1},\dots,i_{d}}\vert\psi_{i_{1}}\dots\psi_{i_{d(k)}})$, with $d(k) = k(q-2) + p$ for a $p$-body initial operator. Using this expression, we find the time evolved $\mathfrak{q}$-complexity as follows
\begin{align}
    \mathcal{Q}(t) = \sum_{s, t, k, l}\,\sum_{i_{1},\dots,i_{s},j_{1},\dots,j_{t}}i^{k - l}\,\varphi^{*}_{l}(t)\,q^{i_{1},\dots,i_{s}}_{j_{1},\dots,j_{t}}\,\varphi_{k}(t)(\mathcal{O}_{l}\vert \psi_{i_1}\dots\psi_{i_s})(\psi_{j_1}\dots\psi_{j_t}\vert \mathcal{O}_{k})\,.
\end{align}
The inner products in the above expression are evaluated by assuming that all the operators are properly normalized (or factors absorbed in $ $) and using the fact that strings of different lengths are orthogonal. This simplifies the expression to the following
\begin{align}
    \mathcal{Q}(t) = \sum_{k, l}\,\sum_{i_{1},\dots,i_{d(l)},j_{1},\dots,j_{d(k)}
    }i^{k - l}\,\varphi^{*}_{l}(t)\,q^{i_{1},\dots,i_{d(l)}}_{j_{1},\dots,j_{d(k)}}\,\varphi_{k}\,(t)\,c^{*}_{i_{1},\dots,i_{d(l)}}c_{j_{1},\dots,j_{d(k)}}\,.
\end{align}
For $q$ even, the string length $d(k)$ can be either odd or even for a fixed $p$ for all $k$. This is as far as one can go in general. However, using the disordered nature of our system, we can make the reasonable assumption that (with small variance and zero mean) under disorder averaging one can write
$\langle c^{*}_{i_{1},\dots,i_{d(l)}}c_{i_{1},\dots,i_{d(k)}} \rangle = \delta_{i_{1},j_{1}}\dots\delta_{i_{d(l)},i_{d(k)}}\delta_{k, l} 2^{d(l)} \binom{N}{d(l)}^{-1}$. This greatly simplifies the expression for the (disorder-averaged) $\mathcal{Q}(t)$ further, leading to
\begin{align}
    \langle\mathcal{Q} (t)\rangle = \sum_{l}2^{l} \binom{N}{d(l)}^{-1}  \vert\varphi_{l}(t)\vert^2\sum_{i_{1},\dots,i_{d(l)}} q^{i_{1},\dots,i_{d(l)}}_{i_{1},\dots,i_{d(l)}}\,.\label{qcomp}
\end{align}
Further computation would require exact knowledge of the coefficients $q^{i_{1},\dots,i_{d(l)}}_{i_{1},\dots,i_{d(l)}}$. Specific choices of these coefficients lead us to the expressions for the K-complexity, OTOC, operator size, etc. For these three probes the exact coefficients are discussed in \cite{Parker:2018yvk}. 
%\subsection{For random cubic jump operators}

%\textcolor{red}{We need to fill this section}

\section{Summary and conclusions} \label{conclusion}
In this paper, we performed a detailed study of operator growth through Krylov complexity in Lindbladian SYK, where the dissipation is modeled by various jump operators in the Markovian regime. In particular, we choose our system to be SYK and model the dissipation by $p$-body Lindblad operators. We analytically find the Lanczos coefficients which are numerically verified by implementing the bi-Lanczos algorithm, a suitable generalization of the Lanczos algorithm in open systems. We obtained a universal result of Krylov complexity, which initially grows exponentially and saturates at late times. Both the saturation time scale and saturation value appear to be generic and independent of the choice of the Lindblad operators, similar to what we obtained from OTOC. We also provide a plausible explanation of our results from the quantum measurement perspective. Based on these findings and analyzing the generic pole structure of auto-correlation and high-frequency behavior of the spectral function, we propose an operator growth hypothesis for generic open quantum systems, which suggests a broader notion of ``dissipative quantum chaos''.

Our approach opens a door to understanding a generic study of operator growth and chaos in non-Hermitian systems. Especially, since any Hermitian matrix can always be tri-diagonalizable and put into the form \eqref{ld} with $a_n = 0$, we wonder what could be a Hamiltonian structure of \eqref{ld}. In particular, for the Lindblad evolution, the density matrix evolves as \cite{Matsoukas-Roubeas:2022odk}
\begin{align*}
    \dot{\rho} = - i \big(H_{\mathrm{eff}} \rho - \rho H_{\mathrm{eff}}^{\dagger} \big)\,, ~~~~~ H_{\mathrm{eff}} = H - \frac{i}{2} \sum_m L_m^{\dagger} L_m \,,
\end{align*}
where $H_{\mathrm{eff}}$ is known as the \emph{effective} Hamiltonian which is non-Hermitian.\footnote{Note that it is generically non-Hermitian i.e., $H_{\mathrm{eff}}^{\dagger} \neq H_{\mathrm{eff}}$. It is not anti-Hermitian.} This non-Hermitian Hamiltonian is constructed by the jump operators.\footnote{Regardless of the behavior of the jump operator under Hermitian conjugation. In the main text, we restrict ourselves to Hermitian or anti-Hermitian jump operators only.} It is important to note that the first term $H$ is Hermitian while the second term is anti-Hermitian, which makes the overall Hamiltonian $H_{\mathrm{eff}}$ non-Hermitian. Hence, we wonder whether any non-Hermitian Hamiltonian (including PT-symmetric \cite{Garcia-Garcia:2021elz, Garcia-Garcia:2022xsh}) can be cast into the structure of \eqref{ld}, which can be obtained by an efficient implementation of the bi-Lanczos algorithm.

A limitation of our analysis is that we are completely blind to the physics with stronger coupling. A more broad analysis is to consider a generic coupling, not necessarily a small one.  In fact, some preliminary analysis shows that the Lanczos coefficients become unstable at stronger coupling, which results from the fact that the Markovian approximation breaks down. It will be interesting to explore the non-Markovian regime where a generic coupling can be chosen. In fact, detailed knowledge about the environment (which might be another SYK) might lead to the study of the system in a purely scrambling or dissipative phase \cite{Zhang:2022knu}, resulting in an environment-induced phase transition. Finally, our bigger aim is to develop a systematic and coherent picture to understand dissipative quantum chaos in holographic duality. In particular, since the dual picture of generic $p$-body dissipator is unknown, it is interesting to see if our study in open SYK leads to the appreciation of a clearer picture of de Sitter space in quantum gravity. We hope to address this question in future studies.
\\
\newline
\emph{Note added:} During the final stages of this work, Ref.\,\cite{NSSrivatsa:2023qlh} appeared which deals with the behavior of the high-frequency regime of the auto-correlation. 
Their choice of auto-correlation function, specific types of system, and dissipation is fundamentally different from ours. Hence, we do not make any comparison with our results with Ref.\,\cite{NSSrivatsa:2023qlh}.

\section*{Acknowledgements}

We would like to thank Alexei Andreanov, Aranya Bhattacharya, Xiangyu Cao, Saskia Demulder, Anatoly Dymarksy, Dami\'an A. Galante, Hosho Katsura, Tokiro Numasawa, Dario Rosa, Shinsei Ryu, Masaki Tezuka, and Hironobu Yoshida for discussions on related topics. Numerical calculations were performed in the workstation Octaquark-4, CHEP, and using the computational facilities of YITP. P.N. acknowledges the hospitality of the Indian Institute of Science, Saha Institute of Nuclear Physics, The University of Tokyo during the final stages of the work and the long-term workshop YITP-T-23-01 held at YITP, Kyoto University, where some of the results were presented. T.P. thanks the generous hospitality of the Department of Physics, Kyoto University. B.B. acknowledges the financial support from the Institute for Basic Science (IBS) through the project IBS-R024-D1.  The work of P.N. is supported by the JSPS Grant-in-Aid for Transformative Research Areas (A) ``Extreme Universe'' No.\,21H05190. 

\appendix

\section*{Appendices}

\section{Random quadratic Lindbladian} \label{randquadproof}

In this Appendix, we provide a derivation of \eqref{prop32}. The generic expression to be evaluated is the following
\begin{align}
    \mathcal{L}_{D}^{\dagger}\,\mathcal{O} = -i \sum_{k = 1}^{M}\Big[ &\left(\sum_{1 \leq i < j \leq N}V^{k}_{i j}\psi_{i}\psi_{j} \right)^\dagger \mathcal{O} \left(\sum_{1 \leq p < q \leq N}V^{k}_{p q}\psi_{p}\psi_{q} \right) \nonumber \\
    &- \frac{1}{2}\Big\{ \left(\sum_{1 \leq i < j \leq N}V^{k}_{i j}\psi_{i}\psi_{j} \right)^\dagger  \left(\sum_{1 \leq p < q \leq N}V^{k}_{p q}\psi_{p}\psi_{q} \right), \mathcal{O}\Big\}\Big]\notag\,.
\end{align}
We note that $\left(V^{k}_{i j}\psi_{i}\psi_{j}\right)^\dagger = -\overline{V}^{k}_{i j}\psi_{i}\psi_{j}$. Thus, we can write the following simpler expression
\begin{align}
    \mathcal{L}_{D}^{\dagger}\,\mathcal{O} = i \sum_{k = 1}^{M}\sum_{1 \leq i < j \leq N}\sum_{1 \leq p < q \leq N}\overline{V}^{k}_{i j}V^{k}_{p q}\Big[\psi_{i}\psi_{j} \mathcal{O} \psi_{p}\psi_{q} - \frac{1}{2}\Big\{ \psi_{i}\psi_{j}\psi_{p}\psi_{q}, \mathcal{O}\Big\}\Big]\,. \label{LdO}
\end{align}
Since our goal is to derive the action of the dissipative part of the Lindbladian on $\mathcal{O}_n$, the strategy is then to look at the term in the square brackets with $\mathcal{O} \equiv \mathcal{O}_n$ given by \eqref{Onterm}. We write this term as follows
\begin{align}
    \mathcal{A}^{i_{1},\dots,i_{s}}_{(i,j)(p,q)} = \psi_{i}\psi_{j} (\psi_{i_{1}}\psi_{i_{2}}\cdots \psi_{i_{s}}) \psi_{p}\psi_{q} - \frac{1}{2}\Big\{ \psi_{i}\psi_{j}\psi_{p}\psi_{q}, \psi_{i_{1}}\psi_{i_{2}}\cdots \psi_{i_{s}}\Big\}\,. \label{Aij}
\end{align}
To further simplify parts of the calculation, we consider the following relations
\begin{align}
    \{\psi_{j_{1}}\psi_{j_{2}}\cdots\psi_{j_{t}},\psi_{i_{1}}\psi_{i_{2}}\cdots \psi_{i_{s}}\} = (1 + (-1)^{t s})\left(\psi_{i_{1}}\psi_{i_{2}}\cdots \psi_{i_{s}}\right)\left(\psi_{j_{1}}\psi_{j_{2}}\cdots\psi_{j_{t}}\right) \,,\label{AcommRel}\\
    [\psi_{j_{1}}\psi_{j_{2}}\cdots\psi_{j_{t}},\psi_{i_{1}}\psi_{i_{2}}\cdots \psi_{i_{s}}] = (-1 + (-1)^{t s})\left(\psi_{i_{1}}\psi_{i_{2}}\cdots \psi_{i_{s}}\right)\left(\psi_{j_{1}}\psi_{j_{2}}\cdots\psi_{j_{t}}\right)\,, \label{CommRel}
\end{align}
which holds when there are no common indices between the sets $\{j_{1},j_{2},\dots,j_{t}\}$ and $\{i_{1},i_{2},\dots,i_{s}\}$. We denote the first and second terms in \eqref{Aij} as $G_{1}$ and $G_{2}$ respectively. Thus one can write $\mathcal{A}^{i_{1},\dots,i_{s}}_{(i,j)(p,q)} = G_{1} - \frac{1}{2}G_{2}$. We consider the various cases in terms of the indices $\{i, j, p, q\}$ of the summation. 

First, we show that the action of the dissipative part of the Lindbladian to single-string operator $\psi_1$ results in the following expression
\begin{align}
    \mathcal{L}_{D}^{\dagger} \,\psi_{1} = i R V^2\,\psi_{1}\,, \label{prop2}
\end{align}
where $R = M/N$ and $V$ is given by the ensemble average \eqref{vn}. This result is strictly valid in the large $N$ limit.
The proof goes as follows. For the single-string operator $\mathcal{O} = \psi_{1}$, we have to evaluate
\begin{align}
    \mathcal{A}^{1}_{(i,j)(p,q)} = \psi_{i}\psi_{j} \psi_1 \psi_{p}\psi_{q} - \frac{1}{2}\Big\{ \psi_{i}\psi_{j}\psi_{p}\psi_{q}, \psi_1\Big\}\,.
\end{align}
There are naturally $3$ cases where we may expect a non-zero contribution those are for $i = 1, p \neq 1$, $i \neq 1, p = 1$, and $i = 1, p  =  1$. This is because we necessarily have $j > i$ and $q > p$. For these three cases, we obtain the following
\begin{align}
    A^{1}_{(1,j),(p,q)} = -\frac{1}{2}\psi_{j}\psi_{p}\psi_{q}\,, ~~~~
    A^{1}_{(i,j),(1,q)} = \frac{1}{2}\psi_{i}\psi_{j}\psi_{q}\,, ~~~~
    A^{1}_{(1,j),(1,q)} = \psi_{1}\psi_{j}\psi_{q}\,.
\end{align}
Plugging this in \eqref{LdO}, we get the following (suppressing the summation over $k$ for now)
\begin{align}
    -2i(\mathcal{L}^{k}_{D})^{\dagger} \, \psi_{1} = 2\psi_1\sum_{1<j,q\leq N}\overline{V}^{k}_{1j}V^{k}_{1 q}\psi_{j}\psi_{q} &+ \sum_{1 < i < j \leq N}\sum_{1<q\leq N}\overline{V}^{k}_{i j}V^{k}_{1 q}\psi_{i}\psi_{j}\psi_{q} \nonumber \\
    &- \sum_{1 < p < q \leq N}\sum_{1<j,q\leq N}\overline{V}^{k}_{1 j}V^{k}_{p q}\psi_{j}\psi_{p}\psi_{q}\notag\,,
\end{align}
where $\mathcal{L}_{D}^{\dagger} = \sum_k (\mathcal{L}^{k}_{D})^{\dagger}$. A slight renaming and rearranging of the indices gives us the following expression
\begin{align}
    -2i(\mathcal{L}^{k}_{D})^{\dagger} \,\psi_{1} = \left(\sum_{j = 2}^{N}\vert V^{k}_{1 j}\vert^2\right)\psi_1 &+ \sum_{1 < j < q \leq N}\left(\overline{V}^{k}_{1 j}V^{k}_{1 q} -  V^{k}_{1 j}\overline{V}^{k}_{1 q}\right)\psi_j \psi_q \nonumber \\
    &+ \sum_{1 < p < q \leq N}\sum_{1 < j \leq N}\left(\overline{V}^{k}_{p q}V^{k}_{1 j} -  V^{k}_{p q}\overline{V}^{k}_{1 j}\right)\psi_{j}\psi_{p}\psi_{q}\notag\,.
\end{align}
Clearly, this term is not $\propto \psi_{1}$. However, each term in the parenthesis is also a random variable with the same mean and the variance as $V_{ij}$. Thus, under the condition that $V^{k}_{i j}V^{k}_{p q}$, (with $i\neq j$ and/or $p \neq q$) vanishes upon averaging (for small $V$), the second and third terms in the above expression vanish and we are left with
\begin{align}
    \mathcal{L}_{D}^{\dagger} \,\psi_{1} = \frac{i}{2}\left(\sum_{k = 1}^{M}\sum_{j = 2}^{N}\langle\vert V^{k}_{1 j} \vert^2\rangle\right)\psi_{1}\,.
\end{align}
Moreover, we have the result $\langle\vert V^{k}_{1 j} \vert^2\rangle = \frac{2 V^2}{N^2}$. Using $M = R N$, we obtain
\begin{align}
    \mathcal{L}_{D}^{\dagger} \,\psi_{1} = \frac{i}{2} \frac{2 V^2}{N^2} M (N-1) = i RV^2\left(1 - \frac{1}{N}\right)\,\psi_{1}\,. \label{aa}
\end{align}
In the large $N$ limit, this concludes  \eqref{prop2}.\\
\newline
Now we consider a general string of length $s = n(q-2)+1$ and state the following proposition \eqref{prop32}.\\
\newline
\textbf{Proposition 3.} \emph{The action of the dissipative part of the Lindbladian to a string of length $s = n(q-2)+1$ results in the following expression
\begin{align}
    \mathcal{L}_{D}^{\dagger} \, \mathcal{O}_n = i \zeta q R V^2 n \, \mathcal{O}_n\,,
\end{align}
where $\zeta \sim o(1)$ number and $V$ is given by the ensemble average \eqref{vn}.}\\
\newline
\emph{Proof:} Here we discuss the general case for an operator string $\psis$, where $s = n(q-2) + 1$ is an odd number. For this, we note from the expression \eqref{LdO} that the only terms that will survive after averaging over multiple realizations are the ones with $i = p$ and $j = q$. We only evaluate terms of this kind below. There are three such distinct cases.
\subsubsection*{Case 1, $\{i_{m},i_{n},i_{m},i_{n}\}$:}
Here all four indices $\{i_{m},i_{n},i_{m},i_{n}\}\in \{i_{1},i_{2},\cdots,i_{s}\}$ and the two pairs are identical. Here we find 
\begin{align*}
    G_{1} &= \psi_{i_m}\psi_{i_n}(\psis)\psi_{i_m}\psi_{i_n} = \frac{1}{4}(-1)^{2 s - 1}\psis\,, \\
        G_{2} &= -\frac{1}{4}\{I,\psis\} = -\frac{1}{2}\psis\,.
\end{align*}
where for the second equality, we used the fact that $\psi_{i_m}\psi_{i_n}\psi_{i_m}\psi_{i_n} = -\frac{1}{4}I$. From here it follows that $ G_{1} - \frac{1}{2}G_{2} = 0$. Hence, this case does not contribute to \eqref{Aij}.
\subsubsection*{Case 2, $\{i,j,i,j\}$:}
Here both the pairs are identical. Using the identity $\psi_{i}\psi_{j}\psi_{i}\psi_{j} = - \frac{1}{4}I$, we obtain
\begin{align*}
    G_{1} &= \psi_{i}\psi_{j}\psis \psi_{i}\psi_{j} = \frac{1}{4}(-1)^{s+1}(-1)^{s}\psis = -\frac{1}{4}\psis \,, \nonumber \\
      G_{2} &= -\frac{1}{4}\{I,\psis\} = -\frac{1}{2}\psis\,,
\end{align*}
and hence, we find $G_{1} - \frac{1}{2}G_{2} = 0$. This term also has no contribution.
\subsubsection*{Case 3, $\{i_m, j, i_m, j\}$:}
Here two of the identical indices belong to the set $\{i_m\} \in \{i_{1},i_{2},\cdots,i_{s}\}$ but the other two (also identical and $> i_m$) indices do not. Therefore, we note that $\psi_{i_m}\psi_{j}\psi_{i_m}\psi_{j} = -\frac{1}{4}I$. The two terms are respectively
\begin{align*}
    G_{1} &= \psi_{i_m}\psi_{j}\psis\psi_{i_m}\psi_{j} = \frac{1}{4}(-1)^{s + 1}(-1)^{s-1}\psis = \frac{1}{4}\psis\,, \nonumber \\
    G_{2} &= -\frac{1}{4}\{I,\psis\} = -\frac{1}{2}\psis\,.
\end{align*}
Thus we obtain the expression $G_{1} - \frac{1}{2}G_{2} = \frac{1}{2}\psis$
\subsubsection*{Case 4, $\{j, i_m, j, i_m\}$:}
This is almost identical to the previous case, except that now we have $j < i_m$. Again, the terms $G_1$ and $G_2$ can be written as
\begin{align*}
    G_1 &= \psi_j \psi_{i_m}\psis\psi_j\psi_{i_m} = \frac{1}{4}\psis\,, \nonumber \\
    G_2 &= -\frac{1}{4}\{I, \psis\} = -\frac{1}{2}\psis\,.
\end{align*}
From this we obtain the expression $G_1 - \frac{1}{2}G_2 = \frac{1}{2}\psis$.

Keeping these cases in mind, we find the following expression for \eqref{LdO} as
\begin{align}
    \mathcal{L}_{D}^{\dagger}\,\mathcal{O}_{n} = \frac{i}{2}\left(\sum_{k = 1}^{M}\sum_{1 \leq i < j \leq N}\langle\vert V^{k}_{i j}\vert^2\rangle \right)\Big\vert_{i / j \in \{i_1,\dots,i_s\}}\mathcal{O}_{n} = C_n \,\mathcal{O}_{n}\,, \label{lf}
\end{align}
where we have introduced the $\langle \,\, \rangle$ notation to denote the averaging. The notation $i/j \in \{i_1,\dots,i_s\}$ is taken to imply that either $i$ or $j$ lie in $\{i_1,\dots,i_s\}$ but not both. The objective now is to evaluate the coefficient
\begin{align}
    C_{n} = \frac{i}{2}\left(\sum_{k = 1}^{M}\sum_{1 \leq i < j \leq N}\vert V^{k}_{i j}\vert^2 \right)\Big\vert_{i / j \in \{i_1,\dots,i_s\}}\,.
\end{align}
We break this sum over $i, j$ into the following pieces
\begin{align}
    \sum_{1 \leq i < j \leq N}\langle\vert V^{k}_{i j}\vert^2\rangle  = \sum_{i \in \{i_1,\dots,i_s\}}\sum_{j > i}\langle\vert V^{k}_{i j}\vert^2\rangle + \sum_{i < j}\sum_{j \in \{i_1,\dots,i_s\}}\langle\vert V^{k}_{i j}\vert^2\rangle - 2\sum_{i_{m}}\sum_{i_{n} > i_{m}}\langle\vert V^{k}_{i j}\vert^2\rangle\,.
\end{align}
The reason for subtracting the third term is because it is included in both the first and second terms, where we place no constraints on $j$ other than $j > i$. Additionally, under the averaging condition, we can use the result $\langle\vert V^{k}_{i j}\vert^2\rangle = \frac{2 V^2}{N^2}$. Using this, the summation turns out to be the following
\begin{align}
    \sum_{1 \leq i < j \leq N}\langle\vert V^{k}_{i j}\vert^2\rangle &= \frac{2 V^2}{N^2}\left( \sum_{i \in \{i_1,\dots,i_s\}}\sum_{j}1 - \sum_{i \in \{i_1,\dots,i_s\}}1 - \sum_{i_{m}}\sum_{i_{n} > i_{m}}2\right)\,,\notag \\
    &= \frac{2 V^2}{N^2}\left((N-1)s - 2\frac{s(s-1)}{2}\right)\,, \\
    &= \frac{2 V^2}{N} s \left(1 - \frac{s}{N}\right)\,.
\end{align}
Hence the coefficient is
\begin{align}
    C_{n} = \frac{i}{2}\frac{2 V^2}{N} s M \left(1 - \frac{s}{N}\right) = i RV^2 s \left(1 - \frac{s}{N}\right)\,,
\end{align}
where $R = M/N$. For $s=1$, this reduces to \eqref{aa}. However, our interest is in the large $N$ and large $q$ limit, where we approximate $s \approx n q$ for large $q$. Thus, we obtain
\begin{align}
    C_{n} = \frac{i}{2}\frac{2V^2}{N} M n q \left(1 - \frac{n q}{N}\right) \sim \zeta q RV^{2}n\,, \label{cnp2}
\end{align}
where $\zeta \sim o(1)$ is a number independent of $n$ and $V$. The important conclusion here is that this coefficient $C_{n}$ is proportional to $V^2$, $R$ and $n$. One subtlety to be noted here is that this analysis holds for an operator string $\psis$ that is long enough (i.e $s < N$), which is evident in the large $q$ approximation.  Plugging Eq.\,\eqref{cnp2} in Eq.\,\eqref{lf}, we obtain Eq.\,\eqref{prop32}.

\section{Random $p$-body Lindbladian}  \label{randpbodyproof}

To derive \eqref{prop4}, we consider the general Lindblad equation
\begin{align}
    \mathcal{L}_o^{\dagger} \,\mathcal{O} = [H,\mathcal{O}] - i \sum_{k = 1}^{M}[(-1)^{p s}L^\dagger_{k}\mathcal{O}L_{k} - \frac{1}{2}\{L^\dagger_{k}L_{k},\mathcal{O}\}]\,,
\end{align}
where $p$ is the number of fermions in $L_{k}$ and $s$ is the number of fermions in $\mathcal{O}$. We represent the operators $L_{k}$ and $\mathcal{O}$ by the following expressions
\begin{align}
    L_{k} &= \sum_{1\leq \alpha_1 < \alpha_2 < \dots < \alpha_{p} \leq N}\val\psial\,, \\
    \mathcal{O} &= \psis\,.
\end{align}
For compactness, we denote the sum $\sum_{1\leq \alpha_1 < \alpha_2 < \dots < \alpha_{p}}$ ans $\sum_{\{\alpha, p\}}$ in what follows. The dissipative part of the Lindbladian will be the central focus of our analysis. This is written as
\begin{align}
    \mathcal{L}_{D}^{\dagger} \,\mathcal{O} = &-i(-1)^{p(p-1)/2}\sum_{k = 1}^{M}\Big[(-1)^{p s}\left(\sum_{\{\alpha, p\}} \ovral\psial\right)\mathcal{O}\left(\sum_{\{\beta, p\}} \vbe\psibe\right) \notag \\
    &-\frac{1}{2}\sum_{\{\alpha,p\}}\sum_{\{\beta,p\}}\Big\{ \ovral \psial \vbe \psibe, \mathcal{O}\Big\}
    \Big]\,,
\end{align}
where we have used the following fact
\begin{align}
    L^\dagger_{k} = (-1)^{p(p-1)/2}\sum_{\{\alpha,p\}}\ovral \psial\,.
\end{align}
We emphasize that the coefficients $\val$ are taken from a random complex Gaussian distribution with zero mean and variance $\langle \vert \val \vert^2 \rangle = \frac{p! V^2}{N^p}$. This implies that for small $V$ and large $N$, we can ignore terms of the kind $\ovral\vbe$ with some indices different between the sets $\{\alpha_p\}$ and $\{\beta_p\}$ since these will vanish upon averaging over a large number of realizations. We focus our attention to terms with $\alpha_i = \beta_i\,\forall\, i$. With these terms only, the average over Lindbladian may be written as
\begin{align}
    \mathcal{L}_{D}^{\dagger} \,\mathcal{O} = -i(-1)^{p(p-1)/2} \sum_{k = 1}^{M}\sum_{\{\alpha,p\}}\langle\vert \val \vert^2\rangle &\Big[ \psial \mathcal{O} \psial(-1)^{p s} 
    \notag\\ &- \frac{1}{2}\{\psial \psial,\mathcal{O}\}\Big]\,.
\end{align}
The terms in the squared parenthesis need to be treated with care. We split these into two terms
\begin{align}
    G_1 &= (-1)^{p s}\psial \left(\psis\right) \psial\,,\\
    G_2 &= -\frac{1}{2}\{\psial \psial,\psis\}\,,
\end{align}
where we have used the expression for $\mathcal{O}$ which follows from operator concentration. Firstly, we note that 
\begin{align}
    \psial \psial = \frac{(-1)^{p(p-1)/2}}{2^p}I\,,
\end{align}
which implies that 
\begin{align}
    G_2 = -\frac{(-1)^{p(p-1)/2}}{2^p}\psis\,.
\end{align}
The evaluation of $G_1$ is more involved. For this, we note that any $\alpha_k$ can be written as 
\begin{align}
    \psi_{\alpha_k}\psis = (-1)^{s + \zeta^{s}_l}\psis\psi_{\alpha_{k}}\,, \;\;\;\;\;\text{where}\;\;\; \zeta^{s}_{k} = \sum_{j = 1}^{s}\delta_{i_{j},\alpha_{k}}\,.
\end{align}
Systematically moving each $\alpha_{i}$ across $\psis$ from the right to left (or vice-versa) we pick up a phase of $(-1)^{p - i}$ by the time it reaches $\psis$. At which point it crosses over to the other side picking the phase $(-1)^{s + \zeta^{s}_{i}}$. Then it combines with the corresponding operator on the other side of $\psis$ to give a factor of $\frac{1}{2}I$. Repeating this process for all the $p$ fermions, we find the following result
\begin{align}
    G_1 = \frac{(-1)^{p(p-1)/2}(-1)^{p s}}{2^p}(-1)^{\sum_{l = 1}^{p}\zeta^{s}_l}\psis\,.
\end{align}
Combining $G_1$ and $G_2$, we get the following
\begin{align}
    G_1 + G_2 = \frac{(-1)^{p(p-1)/2}}{2^p}\left((-1)^{\sum_{l = 1}^{p}\zeta^{s}_{l}} - 1 \right)\psis\,.
\end{align}
With this, one can write the averaged dissipative Lindbladian as follows
\begin{align}
    \mathcal{L}_{D}^{\dagger} \,\psis = \frac{i}{2^p} \left(\sum_{k = 1}^{M}\sum_{\{\alpha,p\}}\langle \vert \val \vert  \rangle^2\left(1 - (-1)^{\sum_{l = 1}^{p}\zeta^{s}_{l}}\right)\right)\psis\,.
    \label{LDfullExp}
\end{align}
One feature that we note from the onset is that if none of the $\alpha_{i}$ indices lie in $\{i_1,i_2,\dots,i_2\}$, then we get zero contribution since all the $\zeta^s_{l}$ are vanishing. This was also observed explicitly in the $2$-body jump operator case studied in the previous section. Additionally, it is also evident that there will be a non-zero contribution only when an \textit{odd} number of the indices $\{\alpha_{i}\}$ lie in $\{i_1,\dots,i_s\}$. This was also observed in the $2$-body jump operator case.

The objective now is to perform the combinatorial sum
\begin{align}\label{combsum}
    \frac{1}{2}\sum_{1\leq \alpha_1 < \alpha_2 <\cdots < \alpha_p \leq N}\left(1-(-1)^{\sum_{l = 1}^{p}\zeta^{s}_{l}}\right) =& \left(\sum_{1\leq \alpha_1 < \alpha_2 <\cdots < \alpha_p \leq N}1\right) \notag\\&- \left(\sum_{1\leq \alpha_1 < \alpha_2 <\cdots < \alpha_p \leq N \& \alpha_{i} \notin \{i_1,\dots,i_s\}}1\right) \nonumber \\&-  \left(\sum_{1\leq \alpha_1 < \alpha_2 <\cdots < \alpha_p \leq N \& \alpha_{i, \text{even}} \in \{i_1,\dots,i_s\}}1\right)\,.
\end{align}
It is useful to consider this sum in pieces.  This sum tells us to put $2$ for every case where $1\leq\alpha_1<\cdots<\alpha_p \leq N$ and then subtract $2$ for each case where $\forall \alpha_{i} \notin \{i_1,\dots,i_s\}$. Then the cases where an even number of $\alpha_{i} \in \{i_1,\dots,i_s\}$ also have to correspond to a subtraction of $2$. We denote the L.H.S of the above summation as $\frac{1}{2}(S_1 - S_2 - S_3)$. To evaluate the first step, we note the following sum
\begin{align}
    S_{1} = \sum_{1\leq \alpha_1 < \alpha_2 <\cdots < \alpha_p \leq N} 2 = \frac{2\Gamma(N+1)}{\Gamma(p + 1)\Gamma(N - p + 1)}\,. \label{sum1}
\end{align}
The next step is to subtract out the cases where $\forall \alpha_{i} \notin \{i_1,\dots,i_s\}$. To do this, it is enough to note that the sum will be the same as \eqref{sum1}, just with $N$ replaced by $N - s$. Therefore this term is 
\begin{align}
    S_{2} = \sum_{1\leq \alpha_1 < \alpha_2 <\cdots < \alpha_p \leq N \& \alpha_{i} \notin \{i_1,\dots,i_s\}} 2 = \frac{2\Gamma(N-s+1)}{\Gamma(p + 1)\Gamma(N - p - s + 1)}\,.
\end{align}
Let us now look at the term $S_{1} - S_{2}$. In the large $N$ limit, this terms is given by 
\begin{align}
    S_{1} - S_{2} = \frac{2\Gamma(N+1)}{\Gamma(p + 1)\Gamma(N - p + 1)} - \frac{2\Gamma(N-s + 1)}{\Gamma(p + 1)\Gamma(N - p - s + 1)} \xrightarrow[]{N\rightarrow \infty} \frac{2 p N^{p-1}s}{\Gamma(p + 1)}\,.
\end{align}
The final step is to evaluate the sum where an even number of $\alpha_{i}$ are taken from $\{i_{1},\dots,i_{s}\}$. Since the numbering of the fermions is really arbitrary, it suffices to compute the expression for the arrangement $\{i_{1},\dots,i_{s}\} = \{1,2,\dots,s\}$. Any other arrangement of the indices will give the same result. Since we evaluate the sum for the case where $\{i_{1},\dots,i_{s}\}$ are sequential, the combinatorial factors will arise from the choice of an even number of $\{i_{1},\dots,i_{s}\}$. There will be no combinatorial factors from assigning these indices to $\{\alpha_1,\dots,\alpha_p\}$, since these will necessarily be assigned in an increasing order starting from $\alpha_1$. 

The number of ways of selecting $2 k$ number of indices from the $s$ indices available to us is $\binom{s}{2k}$. Once these $2 k$ elements are chosen and assigned to $\alpha_{1},\dots,\alpha_{2 k}$ in ascending order, the sum can be performed over the remaining $p - 2 k$ summation indices spanning over $N - s$ values. Note that $2 k$ is limited by the $s$ or $s-1$ (depending on $s$ odd or even) or $p$ ($p-1$ if $p$ is odd). This simply amounts to replacing $N$ by $N - s$ and $p$ by $p - 2 k$ in \eqref{sum1}. The resulting summation is
\begin{align}
    \mathcal{S}_{k} = \frac{\Gamma(N - s + 1)}{\Gamma(p-2k + 1)\Gamma(N - s - p + 2k + 1)}\binom{s}{2 k}\,.
\end{align}
The full contribution is then simply $S_{3} = 2\sum_{k = 1}^{\min(\lfloor s/2 \rfloor, \lfloor p/2 \rfloor)}\mathcal{S}_{k}$. The full sum is therefore represented as $S_{1} - S_{2} - S_{3}$, using \eqref{LDfullExp},
\begin{align}
     \mathcal{L}_{D}^{\dagger} \,\psis &= \frac{i R V^{2}p!}{2^{p-1} N^{p - 1}}\Big(\frac{\Gamma(N+1)}{\Gamma(p + 1)\Gamma(N - p + 1)} - \frac{\Gamma(N-s + 1)}{\Gamma(p + 1)\Gamma(N - p - s + 1)}\notag\\ &- \sum_{k = 1}^{\min(\lfloor s/2 \rfloor, \lfloor p/2 \rfloor)}\frac{\Gamma(N - s + 1)}{\Gamma(p-2k + 1)\Gamma(N - s - p + 2k + 1)}\binom{s}{2 k}\Big)\psis\,. \label{LDfinal222}
\end{align}
In the large $N$ approximation, we ignore $S_{3}$ as it is easy to see that the leading order contribution of $\mathcal{S}_{k}$ is $O(N^{p - 2k})$, which is at least one order less than $N^{p - 1}$ which is the leading order contribution from $S_{1} - S_{2}$. Hence,
\begin{align}
    \mathcal{L}_{D}^{\dagger} \,\psis = i\frac{p s}{2^{p-1}} R V^{2}\,
 \psis\,.\label{LDfullExp1}
\end{align}
It is easy to see that this reduces to the leading order contribution \eqref{cnp2} for $p  = 2$. Hence, the ``operator size concentration'' \eqref{Onterm} leads to \eqref{prop4}.

\section{General Initial operator}
It has been shown that operator concentration is a property of the large$-N$, large $q$ SYK model \cite{Bhattacharjee:2022lzy}. Using half-melon diagrams to represent the operators generated by subsequent application of the closed SYK Lindbladian, starting with initial operator $\psi_1$, it was shown that the Lanczos coefficients are given by $b_{1} = \mathcal{J}\sqrt{2/q}$, $b_{n > 1} = \mathcal{J}\sqrt{n(n-1)} $. It is straightforward to derive the same results for a general $p-$body initial operator $\psi_{i_1}\psi_{i_2}\dots\psi_{i_p}$. Before presenting the arguments, let us briefly review the salient features of the construction in \cite{Bhattacharjee:2022lzy}.

The SYK Louivillian $\mathcal{L}_{H}$ can be split into two parts, each corresponding to a forward and backward movement in the Krylov basis respectively
\begin{align}
    \mathcal{L}_{H} = \mathcal{L}_{+} + \mathcal{L}_{-}\,.
\end{align}
Considering an operator basis generated by the action of $\mathcal{L}_{+}$ on the initial operator, it is possible to demonstrate that the basis is orthonormal. This is due to the fact that $\mathcal{L}_{+}$ generates Majorana strings of length larger than the one on which it acts. The largest such string is one that arises when a Majorana fermion (in the operator on which $\mathcal{L}_{+}$ is acting) is replaced by a $(q-1)$ body string of fermions. A diagrammatic approach is discussed in detail in \cite{Bhattacharjee:2022lzy}.

%(\textcolor{red}{Insert diagram here})

Subsequent levels of $\mathcal{L}^{n}_{+}\psi_{1}$ give rise to a rapidly increasing number of diagrams, each of the same length. The red curve is the initial operator and the further black curves represent the subsequent additional operators generated. The coefficients of each diagram are essentially the number of ways it can be constructed out of its parent diagram in the previous level. Careful counting results in the following identity
\begin{align}
    \mathcal{L}_{-}\mathcal{L}^{n + 1}_{+}\psi_1 = \frac{1}{2}n(n+1)\mathcal{L}^{n}_{+}\psi_1\,.
\end{align}
Using this, one can argue that the Krylov basis up to normalization is given by 
\begin{align}
    \mathcal{O}_{n} \propto \mathcal{L}^{+}_{n}\psi_{1}\,.
\end{align}
And the usual action of the Liouvillian on the basis element follows accordingly (for $n > 1$)
\begin{align}
\mathcal{L}_{H}\mathcal{L}^{n}_{+}\psi_{1} = \mathcal{L}^{n + 1}_{+}\psi_{1} + \frac{1}{2}n(n-1)\mathcal{L}^{n-1}\psi_{1}\,,
\end{align}
with the edge cases $n = 0, 1$ represented as
\begin{align}
    \mathcal{L}_H \psi_1 = \mathcal{L}_+ \psi_1 \,,\,~~ \mathcal{L}_H \mathcal{L}_+ \psi_1  = \mathcal{L}_+^{2} \psi_1 + \frac{1}{q}  \psi_1\,.
\end{align}
The edge cases have to be calculated by explicit calculation as follows. 
We first consider the case $n=0$, remember we start with the normalized initial operator $\mathcal{O}_{0} = \sqrt{2} \psi_{1}$
\begin{align}
    \mathcal{L}_{H}(\sqrt{2}\psi_{1}) = \sqrt{2}[H, \psi_{1}]\,.
\end{align}
The SYK Hamiltonian is 
\begin{align}
  H=  \sum_{1 \leq i_{1} < \cdots < i_{q} \leq N} J_{i_{1}\cdots i_{q}} \psi_{i_{1}} \cdots \psi_{i_{q}}\,.
\end{align}
Substituting we get, we omit the summation as it is understood
\begin{equation}\label{eq:ope1}
   \mathcal{L}_{H} \psi_{1} = \sqrt{2} J_{i_{1}\cdots i_{q}} [\psi_{i_{1}} \cdots \psi_{i_{q}}, \psi_{1}] \,.
\end{equation}
The non-zero contribution comes only when $i_{1}= 1$. Using this we get 
\begin{align}\label{eq:o1}
    \mathcal{L}_{H} \psi_{1}  = \sqrt{2} J_{1\cdots i_{q}} [\psi_{1} \cdots \psi_{i_{q}}, \psi_{1}] 
    = -\sqrt{2} J_{1\cdots i_{q}} \psi_{i_{2}} \cdots \psi_{i_{q}}\,.
\end{align}
For the $n=1$ case we have, using the action of $\mathcal{L}_{+}$ and $\mathcal{L}_{-}$ \cite{Caputa:2021sib}
\begin{align}
    \mathcal{L}_+ \psi_1  = \mathcal{L}_+^{2} \psi_1 + b_{1}^{2}  \psi_1\,.
\end{align}
We know from the definition that $b_{1}^{2} = ||\mathcal{L}_{H} (\sqrt{2}\psi_{1}) ||$. We can calculate this norm as follows
\begin{align}
 b_{1}^{2}=2 ||\mathcal{L}_{H} \psi_{1} || = 2\frac{\text{Tr}((\mathcal{L}_{H} \psi_{1})^{\dagger}\mathcal{L}_{H} \psi_{1})}{\text{Tr}(\mathrm{I})} \,. 
\end{align}
Using Eq.\,\eqref{eq:o1} and writing the explicit sum( notice the sum), we then have
\begin{equation}\label{equ:b1sq}
   b_{1}^{2} =2\sum_{1 < i_{2} < \cdots < i_{q} \leq N} |J_{1\cdots i_{q}}|^{2} \frac{1}{2^{q-1}} \,.
\end{equation}
Here, we would have to do a disordered averaging. We further use the fact that  $|<J_{1\cdots i_{q}}>|^{2} = \frac{(q-1)^{2} J^{2}}{N^{q-1}}$ 
\begin{equation*}
    \sum_{1 < i_{2} < \cdots < i_{q} \leq N}( 1 )= \frac{\Gamma(N)}{\Gamma(q)\Gamma(N-q+1)}\,.
\end{equation*}
With this, we now have 
\begin{equation}
    b_{1}^{2} = \frac{\Gamma(N)}{\Gamma(q)\Gamma(N-q+1)} \frac{(q-1)! J^{2}}{N^{q-1}} \frac{1}{2^{q-1}} \xrightarrow[]{N\rightarrow \infty} 2^{2-q} J\,.
\end{equation}
Using the fact that in the large $q$ limit $2^{2-q} J ^{2} = 2\mathcal{J}^{2}/q$
\begin{equation}
    b_{1}^{2} = \frac{2\mathcal{J}^{2}}{q}\,.
\end{equation}
With this we can read-off the Lanczos coefficients $b_n = \sqrt{n (n-1) / 2}$ for $n> 1$ and $b_1 = \sqrt{1/q}$. Therefore the Krylov vectors can be generated by subsequent application of $\mathcal{L}_{+}$ on the initial vectors and the corresponding Lanczos coefficients can be easily determined. The length of the leading order Majorana string for a given $n$ is $s = n(q-2) + 1$.

All the arguments discussed here can be extended to a general $p-$ body initial operator. To see this, note that a general $p-$body operator can be represented as a half-melon diagram of size $p$ \eqref{eq:LH1}. 
\begin{align}
    &\psi_{i_1}\psi_{i_2}\dots\psi_{i_p}  \propto  \includegraphics[scale=.5,valign=c]{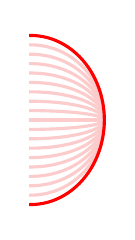} = \includegraphics[scale=.5,valign=c]{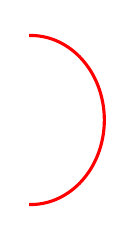} \,. \label{eq:LH1}\\
    &\mathcal{L}_+ \psi_{i_1}\psi_{i_2}\dots\psi_{i_p}  =  c_1\includegraphics[scale=.5,valign=c]{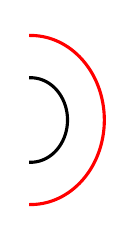} \,, \label{eq:LH2}\\
    &\mathcal{L}_+^2 \psi_{i_1}\psi_{i_2}\dots\psi_{i_p} =  c_2\includegraphics[scale=.5,valign=c]{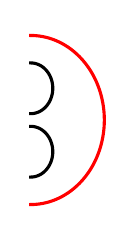} + c_3 \includegraphics[scale=.5,valign=c]{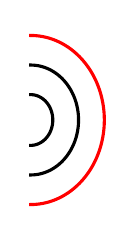}  \,, \label{eq:LH3}\\
    &\mathcal{L}_+^3 \psi_{i_1}\psi_{i_2}\dots\psi_{i_p} =  c_4\includegraphics[scale=.5,valign=c]{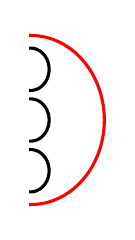} + c_5 \includegraphics[scale=.5,valign=c]{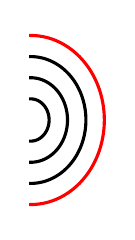} + c_6 \includegraphics[scale=.5,valign=c]{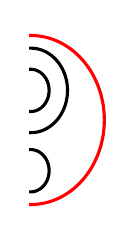} + c_7 \includegraphics[scale=.5,valign=c]{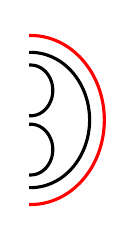}\,.\label{eq:LH4}
\end{align}
%Applying $\mathcal{L}_{+}$ on this diagram consumes one of the fermions and replaces it by $(q - 1)$ fermions \eqref{eq:LH2}. The size of this operator (which is $\mathcal{O}_{1}$ in the Krylov language) is $(q-2) + p$. By definition $\mathcal{L}_{-}$ annihilates the initial operator. To connect it to the discussion of \cite{Bhattacharjee:2022lzy}, it is useful to note that the initial $p-$ body operator can be effectively treated as $\mathcal{L}'_{+}\psi_1$, where $\mathcal{L}'_{+}$ corresponds to a $p-$ body Hamiltonian. Repeating the steps described above, one finds that once again the following relation holds
%\begin{align}
 %   \mathcal{L}_{-}\mathcal{L}^{n + 1}_{+}\psi_{i_1}\psi_{i_2}\dots\psi_{i_p} = \frac{1}{2}(n + 2)(n+1)\mathcal{L}^{n}_{+}\psi_{i_1}\psi_{i_2}\dots\psi_{i_p}
%\end{align}
%This is because of the fact that the removal map of $\mathcal{L}_{-}$ acting on a half-melon diagram is a $n(n+1)/2$-to-$1$ map as before. This allows us to write the expression for the Krylov basis generation as follows
%\begin{align}
 %   \mathcal{L}_{H}\mathcal{L}^{n}_{+}\psi_{i_1}\psi_{i_2}\dots\psi_{i_p} = \mathcal{L}^{n + 1}_{+}\psi_{i_1}\psi_{i_2}\dots\psi_{i_p} + \frac{1}{2}n(n+1)\mathcal{L}^{n-1}\psi_{i_1}\psi_{i_2}\dots\psi_{i_p}
%\end{align}
For the purpose of this manuscript it is enough for us to note that the size of each of these operators at a level $n$ is given by $s = n(q-2) + p$. Additionally, since the Majorana strings of different size are orthogonal, $\mathcal{L}^{n}_{+}\psi_{i_1}\psi_{i_2}\dots\psi_{i_p}$ form a Krylov basis. Since the arguments for the $1-$ body operator should go through (with some minor modifications) here as well, we can expect the asymptotic growth of the Lanczos coefficients to be linear in $n$. Finally, aside from the edge cases of $n = 0, 1$, it is also expected from the half-melon representation that $\mathcal{L}_{-}\mathcal{L}^{n + 1}\psi_{i_1}\psi_{i_2}\dots\psi_{i_p} \propto \mathcal{L}^{n}\psi_{i_1}\psi_{i_2}\dots\psi_{i_p}$. 
The cases for cases $n = 0, 1$ represented as
\begin{align}
    \mathcal{L}_H \psi_{i_1}\psi_{i_2}\dots\psi_{i_p} &= \mathcal{L}_+ \psi_{i_1}\psi_{i_2}\dots\psi_{i_p}\,, \\
    \mathcal{L}_H \mathcal{L}_+ \psi_{i_1}\psi_{i_2}\dots\psi_{i_p}  &= \mathcal{L}_+^{2} \psi_{i_1}\psi_{i_2}\dots\psi_{i_p} + \frac{p}{q}  \psi_{i_1}\psi_{i_2}\dots\psi_{i_p}\,.\label{n1lplus}
\end{align}

%From this, we can read off the expression for the Lanczos coefficients and show that they are identical to the single fermion initial operator case. The Krylov basis is similarly given $\mathcal{O}_{n} \propto \mathcal{L}^{n}_{+} \psi_{i_1}\psi_{i_2}\dots\psi_{i_p}$. The only difference is the length of the operator string, which is $s = n(q-2) + p$. 

We conclude this section by presenting a derivation of \eqref{n1lplus}. To see this, the first step is to note that $\mathcal{L}_{H}\mathcal{L}_{+}\psi_{i_1}\dots\psi_{i_p} = \mathcal{L}^2_{+}\psi_{i_1}\dots\psi_{i_p} + \mathcal{L}_{-}\mathcal{L}_{+}\psi_{i_1}\dots\psi_{i_p}$. In the ladder operator language \cite{Caputa:2021sib}, the second term $\mathcal{L}_{-}\mathcal{L}_{+}\psi_{i_1}\dots\psi_{i_p}$ equals $b_{1}^2\psi_{i_1}\dots\psi_{i_p}$ where $b_1$ is the norm of the $\mathcal{L}_{H}\psi_{i_1}\dots\psi_{i_p}$. We evaluate $b_1$ by starting with the \textit{normalised} initial operator and the SYK$_{q}$ Hamiltonian as given below
\begin{align}
    H  = i^{q/2}\sum_{1 \leq i_{1} < i_{2} < \dots < i_{q} \leq N}J_{i_{1} i_{2} \dots i_{q}}\psi_{i_1}\dots\psi_{i_q}\,, ~~~~~
    \mathcal{O}_{0} = 2^{p/2}\psi_{1}\dots\psi_{p}\,.
\end{align}
Note that we have chosen the indices in our initial operator to be sequential. This is for convenience and does not result in loss of generality. The commutator of $H$ and $\mathcal{O}_{0}$ is given by
\begin{align}
\mathcal{L}_{H}\psi_{i_1}\dots\psi_{i_p} = [H,\mathcal{O}_{0}] = i^{q/2}2^{p/2}\sum_{1 \leq i_{1} < i_{2} < \dots < i_{q} \leq N}J_{i_{1} i_{2} \dots i_{q}}[\psi_{i_1}\dots\psi_{i_q},\psi_{1}\dots\psi_{p}]\,.
\end{align}
The commutator in the sum $[\psi_{i_1}\dots\psi_{i_q},\psi_{1}\dots\psi_{p}]$ evaluates to $(1- (-1)^{l})\psi_{i_1}\dots\psi_{i_q}\psi_{1}\dots\psi_{p}$ where $l$ is the number of the indices in $i_{1},i_{2},\dots,i_{q}$ that coincide in some index from $1,2,\dots,p$. This implies there is a non-zero contribution only when $l$ is odd. The commutator is then denoted as  
\begin{align}
 \mathcal{L}_{H}\psi_{i_1}\dots\psi_{i_p} = 2 i^{q/2}2^{p/2}\sum_{1 \leq \{i,l\} \leq N}J_{i_{1} i_{2} \dots i_{q}}\psi_{i_1}\dots\psi_{i_q}\psi_{1}\dots\psi_{p}\,,
\end{align}
where the index in the summation $\{i,q\}$ indicates the constraint that $l$ of the indices $i_{1},i_{2},\dots,i_{q}$ lie in $1,2,\dots,p$ and there is a sum over all possible $l$. The norm of $[H,\mathcal{O}_0]$ is evaluated to be
\begin{align}
    b_{1}^{2}= \vert\vert\mathcal{L}_{H} \psi_{i_1}\dots\psi_{i_p}\vert\vert = \frac{\text{Tr}((\mathcal{L}_{H} \psi_{i_1}\dots\psi_{i_p} )^{\dagger}\mathcal{L}_{H} \psi_{i_1}\dots\psi_{i_p} )}{\text{Tr}(\mathrm{I})} \,.
\end{align}
This expression evaluates to 
\begin{align}
    b_{1}^{2}= 2^{p + 2}\sum_{1 \leq \{i,l\}\leq N}\vert J_{i_{1} i_{2} \dots i_{q}} \vert^2 \frac{1}{2^{q + p}}\,.
\end{align}
Using the definition of the variance $\langle \vert J_{i_{1} i_{2} \dots i_{q}} \vert^2 \rangle = \frac{(q-1)! J^{2}}{N^{q-1}}$, and the redefinition $2^{1-q}J^2 = \frac{\mathcal{J}^2}{q}$ corresponding to the large$-q$ limit we can simplify the summation in $b_{1}^2$ under disorder averaging to the following expression
\begin{align}
    b_{1}^{2} = 2^{2-q}\frac{(q-1)! 2^{q-1}}{N^{q-1}}\frac{\mathcal{J}^2}{q}\sum_{1 \leq \{i,l\}\leq N}1\,.
\end{align}
This summation has already been evaluated in the previous section \eqref{LDfinal222}. We quote the result here
\begin{align}
    \sum_{1 \leq \{i,l\}\leq N}1 &= \frac{\Gamma(N+1)}{\Gamma(q + 1)\Gamma(N - q + 1)} - \frac{\Gamma(N-p + 1)}{\Gamma(q + 1)\Gamma(N - q - p + 1)}\notag\\ &- \sum_{k = 1}^{\min(\lfloor p/2 \rfloor, \lfloor q/2 \rfloor)}\frac{\Gamma(N - p + 1)}{\Gamma(q-2k + 1)\Gamma(N - p - q + 2k + 1)}\binom{p}{2 k}\,,
\end{align}
In the large $N$ limit, we obtain the following expression
\begin{align}
     b_{1}^{2} = 2^{2-q}\frac{(q-1)! 2^{q-1}}{N^{q-1}}\frac{\mathcal{J}^2}{q}\sum_{1 \leq \{i,l\}\leq N}1 \xrightarrow[N\rightarrow \infty]{} 2\frac{\Gamma(q)}{N^{q-1}}\frac{q N^{q-1}p}{\Gamma(q + 1)}\frac{\mathcal{J}^2}{q} = \frac{2 p \mathcal{J}^{2}}{q}\,.
\end{align}
From this the coefficient $b_{1}$ can be read off as $b_{1} = \mathcal{J}\sqrt{2p/q}$. For our chosen case of $\mathcal{J} =1/\sqrt{2}$, we obtain the result \eqref{n1lplus}.

From this, we can read off the expression for the Lanczos coefficients and show that they are (asymptotically) identical to the single fermion initial operator case. The Krylov basis is similarly given $\mathcal{O}_{n} \propto \mathcal{L}^{n}_{+} \psi_{i_1}\psi_{i_2}\dots\psi_{i_p}$. The only difference is the length of the operator string, which is $s = n(q-2) + p$.

\bibliographystyle{JHEP}
\bibliography{references}

\end{document}